\documentclass[acmsmall,screen]{acmart}

\usepackage{graphicx}
\usepackage{multirow}
\usepackage{booktabs}
\usepackage{amsmath}
\usepackage{makecell}
\usepackage[table]{xcolor}
\usepackage[flushleft]{threeparttable}
\usepackage{listings}
\usepackage{tabularx}
\usepackage[most]{tcolorbox}
\usepackage{xspace}
\usepackage{balance}
\usepackage{soul}
\usepackage{arydshln}

\lstdefinestyle{mystyle}{
    commentstyle=\color{brown},
    keywordstyle=\color{blue},
    numberstyle=\tiny\color{gray},
    stringstyle=\color{orange},
    basicstyle=\ttfamily\footnotesize,
    breaklines=true,                 
    captionpos=b,                    
    numbers=none,                    
    numbersep=5pt,                  
    showspaces=false,                
    showstringspaces=false,
    showtabs=false,                  
    tabsize=2,
    escapeinside={(*@}{@*)},
}
\definecolor{deepgreen}{rgb}{0.0, 0.5, 0.0}
\definecolor{lightgreen}{rgb}{0.7, 1.0, 0.7}
\newtoggle{showChanges}
\togglefalse{showChanges}
\newcommand{\rev}[1]{
  \iftoggle{showChanges}
    {\textcolor{blue}{#1}}
    {#1}
}

\newcommand{\appname}{{\sc MCCom}\xspace}
\newcommand{\appnamebold}{{\sc \textbf{MCCom}}\xspace}

\setcopyright{cc}
\setcctype{by}
\acmDOI{10.1145/3797088}
\acmYear{2026}
\acmJournal{PACMSE}
\acmVolume{3}
\acmNumber{FSE}
\acmArticle{FSE060}
\acmMonth{7}
\received{2025-09-11}
\received[accepted]{2025-12-22}

\begin{document}

\title{Balancing Latency and Accuracy of Code Completion via Local-Cloud Model Cascading}

\author{Hanzhen Lu}
\orcid{0009-0003-6319-2724}
\affiliation{%
  \department{The State Key Laboratory of Blockchain and Data Security and College of Computer Science and Technology}
  \institution{Zhejiang University}
  \city{Hangzhou}
  \country{China}
}
\email{luhanzhen@zju.edu.cn}

\author{Lishui Fan}
\orcid{0009-0006-4602-4296}
\affiliation{%
  \department{The State Key Laboratory of Blockchain and Data Security and College of Computer Science and Technology}
  \institution{Zhejiang University}
  \city{Hangzhou}
  \country{China}
}
\email{12321213@zju.edu.cn}

\author{Jiachi Chen}
\orcid{0000-0002-0192-9992}
\affiliation{%
  \institution{Sun Yat-sen University}
  \city{Zhuhai}
  \country{China}
}
\email{chenjch86@mail.sysu.edu.cn}

\author{Qiuyuan Chen}
\orcid{0000-0002-1240-9095}
\affiliation{%
  \institution{Tencent Technology}
  \city{Shenzhen}
  \country{China}
}
\email{joeqychen@tencent.com}

\author{Zhao Wei}
\orcid{0009-0007-4462-3153}
\affiliation{%
  \institution{Tencent Technology}
  \city{Shenzhen}
  \country{China}
}
\email{zachwei@tencent.com}

\author{Zhongxin Liu}
\orcid{0000-0002-1981-1626}
\authornote{Corresponding author.}
\authornote{Also with Hangzhou High-Tech Zone (Binjiang) Institute of Blockchain and Data Security.}
\affiliation{%
  \department{The State Key Laboratory of Blockchain and Data Security and College of Computer Science and Technology}
  \institution{Zhejiang University}
  \city{Hangzhou}
  \country{China}
}
\email{liu_zx@zju.edu.cn}



\begin{abstract}
Line-level code completion aims to complete the current line in real-time as developers type. 
Low latency is crucial to maintaining a seamless and uninterrupted coding experience, enabling developers to remain in a productive flow. 
However, existing approaches face a fundamental trade-off: large language models (LLMs) provide high-quality suggestions but require expensive computational resources to ensure acceptable inference latency. 
In contrast, static-analysis-based methods and small language models respond quickly but often generate suboptimal completions.
To fill this gap, our idea is to rely on the small model by default and only escalate the large model when necessary to achieve latency-accuracy trade-offs.
Based on this idea, we propose \appnamebold (\textit{M}odel-\textit{C}ascading-based code \textit{Com}pletion), a framework that cascades a local small model with a high-performance cloud large model for code completion.
Realizing effective model cascading requires answering two non-trivial questions, i.e., when to invoke the large model and how to enable effective collaboration between small and large models.
For the first question, we leverage a valuable but easily overlooked signal, i.e., user actions, during code completion to accurately identify failed completions.
This deferral decision allows us to invoke the large model only when necessary, reducing both latency and cloud-side computation costs.
To enable effective collaboration, \appnamebold employs a two-stage speculative decoding strategy and an iterative retrieval mechanism that collectively accelerate and improve the quality of completions.
Due to the lack of high-quality small models for code completion, we also train a lightweight model with only 121M parameters to implement \appnamebold.
The small model achieves an average of 73.8\% of the performance of the state-of-the-art 7B model.
We evaluate \appnamebold on the RepoEval benchmark and a new benchmark, StmtEval, collected from real-world projects. 
Experimental results show that our approach not only reduces inference latency by up to 47.9\% and cuts down LLM usage by an average of 46.3\%, but also improves the exact match rate of the large model by an average of 8.9\%.

\end{abstract}

\maketitle

\section{Introduction}\label{sec:Introduction}
Line-level Code completion refers to the task of completing the current line in real-time as developers type, based on the immediate context of the source code. 
By minimizing the need for manual character-by-character input, code completion significantly improves the efficiency of software development, making it one of the most essential features in modern integrated development environments (IDE)~\cite{amann2016study, murphy2006java}.

The effectiveness of code completion is primarily determined by two key factors: latency and accuracy~\cite{sagtani2025improving}.
In real-world development environments, latency plays a critical role in user experience. 
Low latency ensures seamless interactions, whereas delays in suggestions often lead developers to ignore them and continue coding manually, reducing the tool’s utility.
A recent study shows that 44\% of developers expect statement-level completions within 0.5 seconds~\cite{wangPractitionersExpectationsCode2023}, indicating the importance of responsiveness.
At the same time, high accuracy is equally important. 
Inaccurate or irrelevant suggestions can disrupt the development flow and erode trust in the tool. 
Indeed, 45\% of users identify low-quality completions as the primary issue with current code completion tools~\cite{wangPractitionersExpectationsCode2023}.
These findings highlight the need for code completion systems to strike an effective balance between latency and accuracy for practical adoption.

Current approaches for code completion generally fall into three groups: \textit{static analysis}, \textit{large language models (LLMs)}, and \textit{lightweight local models}.
Static-analysis-based methods mainly rely on static type information, rule-based heuristics, or contextual pattern matching from existing codebases~\cite{bruch2009learning, asaduzzaman2014cscc}.
While effective for single-token completions, these methods have limited capacity in multi-token generation, causing low accuracy.
LLM-based methods have significantly advanced code completion by capturing complex programming patterns and generating high-quality multi-token or even function-level suggestions~\cite{GitHubCopilotYour2025, TabnineAICode, WindsurfFormerlyCodeium, AmazonCodeWhisperer, jiangAiXcoder7BLightweightEffective2024, UniGencoder}.
Nonetheless, due to their large parameter sizes, these models usually require expensive computational resources, are typically deployed in the cloud, and consequently often suffer from high latency. 
Lightweight local models are designed to run entirely on user devices, offering low-latency and responsive completions \cite{semenkin2025full, TongYiLingMa_NiDeZhiNengBianMaZhuShouALiYun}.
However, their limited capacity often results in less effective performance, particularly for complex or ambiguous coding scenarios.
Overall, existing methods struggle to achieve both low latency and high accuracy.

To balance the trade-off between latency and accuracy, we propose a hybrid solution that \textit{cascades a fast, locally-deployed small model with a powerful, cloud-based large model.}
The key idea is to rely on the small model by default and invoke the large model only when necessary, thereby achieving a balance between low latency and high-quality suggestions.

However, realizing effective model cascading presents \textbf{two major challenges}.
\textit{First}, it is non-trivial to determine whether a case can be reliably handled by the local model or requires escalation to the cloud model.
\textit{Second}, designing mechanisms for effective collaboration between the two models remains challenging.
Without such mechanisms, the large model cannot effectively leverage the small model's output, leading to wasted computation and undermining the benefits of cascading.

To address the above challenges, we propose \appname, a \textbf{M}odel-\textbf{C}ascading-based code \textbf{Com}pletion framework that adaptively balances speed and accuracy through behavior-driven routing and collaborative generation.
First, to determine when to escalate to the large model, \appname\ leverages both the local model’s confidence as a proactive indicator and implicit user behavior as a natural feedback signal.
Specifically, during local model generation, we use the confidence of the first three generated tokens to identify and skip samples for which the local model is likely to perform poorly. 
After the local model completes its output, we monitor user behavior, such as continuing to type without accepting the suggestion, as an indicator of dissatisfaction, and trigger a switch to the large cloud-based model if necessary.
Second, to enable effective collaboration between models, \appname\ introduces two mechanisms:

\begin{itemize}
    \item \textit{Two-stage Speculative Decoding}: The system first uses exact-match heuristics to generate a low-cost draft for the small model. 
    If the output of the small model is rejected, it would be reused as a speculative draft for the large model, accelerating decoding at both stages.
    \item \textit{Iterative Retrieval}: Upon rejection, \appname refines retrieval by incorporating the small model's output, enriching the large model’s input and enhancing completion quality.
\end{itemize}

At inference time, given a code snippet to be completed, \appname\ first retrieves a relevant draft from the previous context and invokes the small model to generate a suggestion using speculative decoding.
If the user accepts the suggestion, the process ends.
If rejected, \appname\ retrieves additional semantically relevant context based on the rejected output and invokes the large model, which uses the small model’s output as a speculative draft to produce a refined and more accurate completion.

We evaluate \appname\ with several LLMs on the RepoEval~\cite{zhang2023repocoder} benchmark and our newly constructed benchmark, StmtEval.
RepoEval focuses exclusively on single-line targets, randomly sampling them from real-world repositories to provide a realistic distribution and diversity of coding scenarios.
Different from RepoEval, we argue that line-level code completion should consider a full, functionally complete statement to more accurately reflect the practical utility of code completion tools.
To this end, we introduce StmtEval, which treats a “line” as a complete code statement rather than a single syntactic line.
For instance, in RepoEval, a target can be merely an opening brace \textit{"\{"}, which offers little benefit to the user.
In contrast, StmtEval would instead include the complete statement \textit{"\{\textbackslash n`config.rate': 0.1\textbackslash n\}"}, which provides more useful suggestions.
Moreover, StmtEval contains samples truncated at randomized positions to simulate interactive code completion scenarios.
Together, RepoEval and StmtEval provide a more comprehensive evaluation of line-level code completion systems.
Considering the lack of a high-quality small model for code completion, we further train a lightweight 121M small model from scratch to support fast and efficient completions within our framework.
Despite its compact size, the model achieves an average of 73.8\% of the performance of a state-of-the-art 7B model on benchmark datasets.
Experimental results show that \appname\ consistently delivers strong performance: it achieves 5.8\% to 47.9\% speedup in inference latency, with an average speedup of 25.6\%.
Meanwhile, it outperforms the baseline that always invokes the large model by 2.9\% to 13.5\%, with an average accuracy improvement of 8.9\%, achieving better latency-accuracy trade-offs.
Moreover, \appname\ demonstrates strong generalizability, showing consistent effectiveness across various LLMs.

Our main contributions are as follows.
\begin{itemize}
    \item We propose \appname, a model-cascading framework for code completion that achieves improved latency-accuracy trade-offs. 
    To realize it, we design a behavior-driven routing strategy to decide when to invoke a large model.
    \item We design two collaborative techniques, i.e., two-stage speculative decoding and iterative retrieval, that enable efficient and effective cascading between the local small model and the cloud large model.
    \item We construct a new benchmark that better reflects realistic completion scenarios, addressing the limitations of existing benchmarks in terms of statement granularity and insertion-point randomness.
    \item We perform an extensive evaluation of \appname. 
    Experimental results show that \appname\ improves the performance and accelerates the inference. 
\end{itemize}

\section{Problem Formulation and Motivation}\label{sec:Motivation}
This section formulates the code completion task and describes the motivation for our framework.

\subsection{Problem Formulation}
Each code completion sample can be denoted as $(X_l,X_r,Y,F)$, where $Y$ is the ground-truth completion to be generated, $X_l$ and $X_r$ are the left and right contexts within the same file, and $F$ is the set of other files in the repository.
In practice, the local context $(X_l,X_r)$ may not contain sufficient project-specific information to accurately generate $Y$.
To tackle this problem, state-of-the-art code completion systems usually adopt the retrieval augmented generation (RAG) paradigm.
Specifically, these systems first use a retriever $\mathcal{R}$ to retrieve a set of relevant code snippets $rc_1, \dots, rc_k \subset F$ from the repository based on a query $q$, which is usually constructed based on $X_l$ and/or $X_r$.
The model then conditions on $(X_l, X_r, rc_1, \dots, rc_k)$ to generate the predicted completion $\hat{Y}$.

\subsection{Motivation}

\textit{\textbf{1) Observation 1: A Small Model can Efficiently Handle A Significant Portion of Code Completion Cases.}}
We train a 121M small model with 41M Python line-level code completion and evaluate it on a validation set with 41K samples, where each sample is collected from a Python file in a real-world project (c.f. Section~\ref{sec:training_set}).
The evaluation results show that our small model can produce correct completions for 37.8\% of the cases even without using the RAG paradigm.
Moreover, the 121M model deployed on a client device performs two times faster than a 7B large model deployed on a cloud server with an Nvidia A800 GPU.
As the validation set is collected from real-world repositories, these results imply that a high-quality small model can handle a significant portion of code completion cases with much lower latency.

\noindent{\bf Design.} Based on this observation, if we use the small model to perform completions by default and invoke the large model only when necessary, i.e., cascading them, we can potentially reduce the latency of code completion while maintaining accuracy.
A key challenge, however, lies in deciding when to invoke the large model: calling it too aggressively increases latency, while being too conservative risks lower accuracy.
To address this, we observe that developers naturally express dissatisfaction with suggestions by skipping or deleting them. 
Such interactions provide an implicit yet valuable signal that can guide this decision.

\textit{\textbf{2) Observation 2: Substantial Overlap Among the Local Context and the Outputs of Small and Large Models.}}
We analyze the completions generated by the small model for the 41K samples mentioned above.
We find that 14.4\% of these completions also appear in the left context of the corresponding samples.
In addition, we randomly select 1000 out of the 41K samples, compare the completions of the small and large models for each sample by computing the ES score (c.f. Section~\ref{sec:metric}, which measures the token-level similarity, between them.)
We find that such completions often share some common tokens, and the average ES score is 78.4\%.
These results show that the local context, the output of the small model, and the output of the large model exhibit substantial overlap with one another.

\noindent{\bf Design.} This observation inspires us to further improve the latency of model-cascading-based code generation with speculative decoding.
Speculative decoding generates a draft using a cheaper system and verifies the draft in parallel with the large model to speed up the inference of the large model.
Moreover, the token overlap between the small model’s output and the local context inspires a complementary strategy, i.e., constructing a draft for the small model from the context for another round of speculative decoding.
Based on these thoughts, we introduce a two-stage speculative decoding strategy to speed up the inference of both the small and the large models.  

\begin{table}
\small
\caption{Challenging Task: Combining Models to Improve Accuracy and Speed}
\label{tab:motivation_example}
\begin{threeparttable}
\begin{tabular}{|p{0.95\textwidth}@{\hskip3pt}|}
        \hline
        \textbf{Target code}:\\

        \textcolor{blue}{if}\ self.\_eval\_flag: \\
        \ \ \ \ policy\ =\ create\_policy(self.\_cfg.policy,\ enable\_field=[\textcolor{orange}{'eval'}]).eval\_mode \\
        \textcolor{blue}{else}: \\
        \ \ \ \ policy\ =\ create\_policy(self.\_cfg.policy,\ enable\_field=[\textcolor{orange}{'collect'}]).collect\_mode \\
        self.policy\ =\ policy \\
        \\
        self.\_episode\_result\ =\ [[]\ \textcolor{blue}{for}\ k\ \textcolor{blue}{in}\ \textcolor{blue}{range}(self.\_env\_num)] \\
        \rowcolor{yellow}
        self.\_obs\_pool\ =\ CachePool('obs',\ self.\_env\_num) \\
        \hline
        \textbf{Code retrieved by surrounding lines}: \\
\textcolor{brown}{\#\ create\ policy}\\
\textcolor{blue}{if}\ self.\_eval\_flag:\\
\ \ \ \ assert\ \textcolor{blue}{len}(self.\_cfg.policy)\ ==\ 1\\
\ \ \ \ policy\ =\ [create\_policy(self.\_cfg.policy[0], enable\_field=[\textcolor{orange}{'eval'}])].eval\_mode]\\
\ \ \ \ self.policy\ =\ policy \\
        \hline
        \textbf{Small model:} \\
        \textcolor{deepgreen}{self.\_obs\_pool\ =\ CachePool('obs}\textcolor{red}{\_pool'}\textcolor{deepgreen}{, self.\_env\_num)} \\
        \textbf{Large model:} \\
        \textcolor{deepgreen}{self.\_}\textcolor{red}{episode\_result\_len\ =\ [0 for k in range(}\textcolor{deepgreen}{self.\_env\_num)]} \\
        \hline
        \textbf{Code retrieved by small model's output} \\
        self.\_episode\_result\ =\ [[]\ \textcolor{blue}{for}\ k\ \textcolor{blue}{in}\ \textcolor{blue}{range}(self.\_env\_num)]\\
        self.\_obs\_pool\ =\ CachePool(\textcolor{orange}{'obs'},\ self.\_env\_num)\\
        self.\_policy\_output\_pool\ =\ CachePool(\textcolor{orange}{'policy\_output'},\ self.\_env\_num)\\
        \hline
        \textbf{Large model with iterative retrieval:} \\
        \textcolor{deepgreen}{self.\_obs\_pool = CachePool('obs', self.\_env\_num)} \\
        \hline
    \end{tabular}
    \begin{tablenotes}
        \footnotesize
        \item[1] The code highlighted in yellow indicates the content to be completed.
    \end{tablenotes}
\end{threeparttable}
\end{table}

\textit{\textbf{3) Observation 3: Rejected Completion May Contain Valuable Information.}}
Table~\ref{tab:motivation_example} presents an example from the RepoEval-Line benchmark.
We follow the RAG paradigm and use the small model and the large model with BM25 as the retriever and the left context as the query to generate a completion for this case, respectively. However, neither the small model nor the large model alone can generate the correct completion.
Although the user may reject the incorrect suggestion from the small model, such a suggestion contains useful semantic hints, as it almost predicts the ground truth completion with only one string literal (i.e., ``obs\_pool'') incorrect.
We find that if we use this suggestion as the query to retrieve relevant code snippets from the repository, we can obtain an exact match of the ground truth completion, which could guide the large model to produce the correct result.
This example demonstrates that even failed completions from the small model can serve as valuable anchors for guiding retrieval and further help the large model generate more accurate and contextually grounded completions.

\noindent{\bf Design.} 
Inspired by this observation, we incorporate an iterative retrieval mechanism into \appname,  which leverages the rejected output from the small model to retrieve more relevant code snippets from the repository for better guiding the large model.

\section{Approach}\label{sec:Approach}
\begin{figure*}[t]
    \centering
    \includegraphics[width=0.9\linewidth]{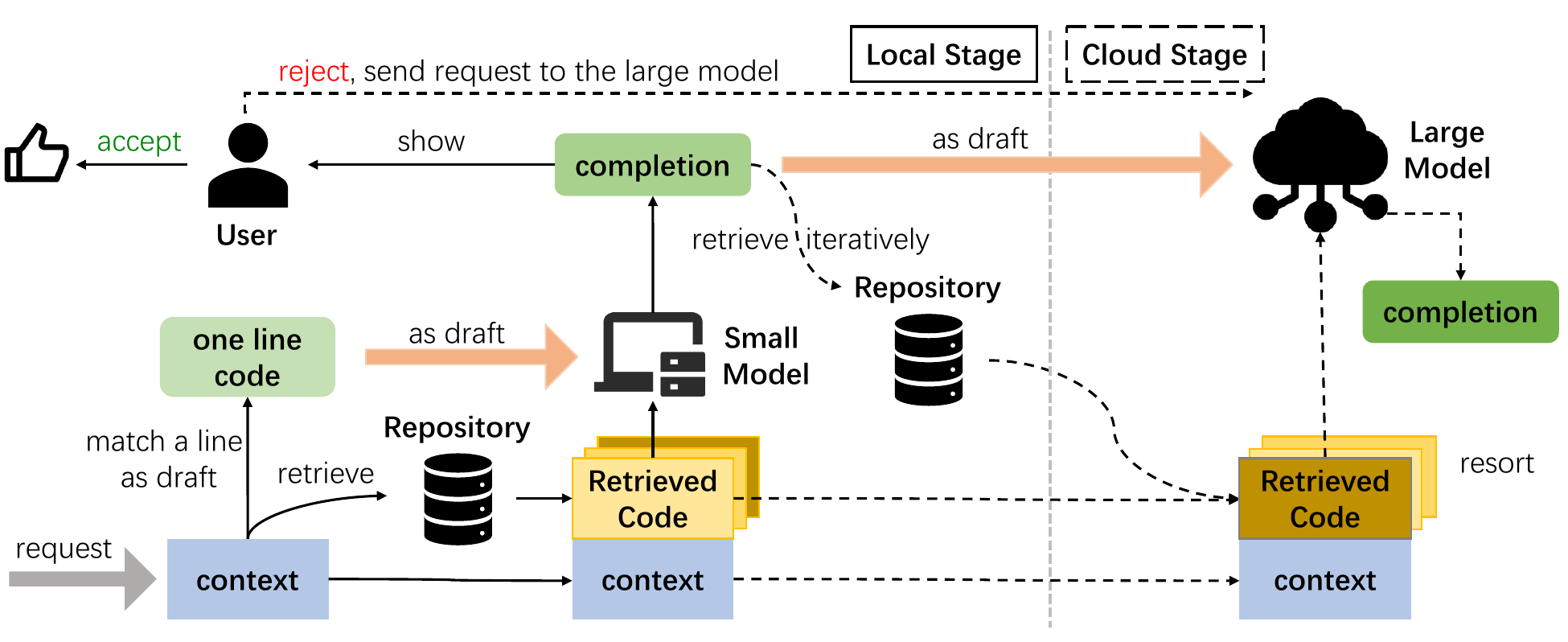}
    \caption{The overview of \appname}
    \label{fig:overview}
\end{figure*}

\subsection{Overview}
Our approach, \appname, aims to combine a locally-deployed small model $\mathcal{SM}$ with a high-capacity cloud-based model $\mathcal{LM}$ to achieve better latency-accuracy trade-offs in code completion. 
Figure~\ref{fig:overview} presents the overall architecture.
To complete a given code snippet, we begin by retrieving relevant code snippets from the repository using the current context. 
In the local stage, we construct a draft by directly matching the line before the completion position against both the context and the retrieved code snippets.
The small model then performs speculative decoding on the draft and generates a completion. 
This candidate completion is presented to the user for fast interaction.
Upon rejection, the system refines retrieval by re-scoring the candidate snippets using the rejected completion, aggregating these scores with the initial retrieval scores, and re-ranking the results. 
The top-ranked snippets are then combined with the in-file context and passed to the large model for further processing.
In the Cloud stage, the large model performs speculative decoding on the output of the small model and generates the final output. 
The rest of this section explains each phase of \appname in more detail.

\subsection{Routing Strategy}
In model cascading, one of the most challenging problems is determining when to invoke a large model. 
This decision is critical because invoking a larger model frequently can result in significant latency, whereas always relying on the small model may hurt the accuracy.

To address this, we introduce a routing strategy composed of two parts.
The first part relies on the confidence of the local model, measured by the probabilities it assigns to its predictions.
\rev{
As reported by prior work~\cite{zhang2025coderag}, log probability can serve as a proxy for the confidence of Code LLMs, which we also corroborate on our validation set. 
Motivated by this, we use the average probability of the first $N$ tokens generated by the local model as the confidence score, denoted as $w_{conf}$. 
Then, we employ a threshold-based routing strategy: if $w_{conf}$ exceeds a predefined threshold, the system accepts the local generation; otherwise, it invokes the cloud model to ensure response quality.
}


The second part of our strategy involves invoking the large model only when necessary, based on implicit user feedback.
When a completion generated by the small model is presented, the system monitors user behavior to interpret feedback implicitly (e.g., by pressing the "Tab" key to accept the completion).
If the user accepts the small model's suggestion, no further action is needed, and the process ends.
However, if the user continues typing, it is treated as an implicit rejection of the suggestion, indicating that the small model’s output was not satisfactory.
This dynamic routing strategy allows the system to balance efficiency and quality by selectively relying on the small model and escalating to the large model only when necessary, minimizing redundant large model calls while ensuring high-quality completions when needed.

\subsection{Two-Stage Speculative Decoding}
To reduce inference latency, we adopt speculative decoding~\cite{leviathan2023fast}, a strategy that allows $\mathcal{LM}$ to validate multiple tokens from the draft in parallel. 
Rather than generating tokens one by one, $\mathcal{LM}$ performs a single forward pass to evaluate the likelihood of a sequence of proposed tokens.
This parallel validation is significantly faster than traditional autoregressive decoding because it amortizes the cost of generating $n$ tokens over a single inference call.
Once speculative validation is complete, all tokens that match the model’s top-1 predictions at their respective positions are considered validated.
The model then resumes generation from the first mismatched token to produce the remaining completion.

\begin{table}[t]
    \centering
    \caption{An example of context-based matching. It operates by:
i) Matching the line preceding the cursor (i.e., \texttt{scheduler = ...}) to a previous occurrence.
2) Retrieving its successor (i.e., \texttt{classifier = ...}) as a draft.}
    \begin{tabular}{|c|} 
        \hline
        \begin{lstlisting}[language=Python, basicstyle=\ttfamily\footnotesize, frame=none, boxpos=c, linewidth=0.9\linewidth]
scheduler = VQDiffusionScheduler(self.num_embed)
classifier = LearnedClassifier(learnable=False)
# ... 
scheduler = VQDiffusionScheduler(self.num_embed)
# <Completion Start>
        \end{lstlisting} \\
        \hline
    \end{tabular}
    \label{tab:context_based_matching}
\end{table}

A straightforward approach is to use the small model to generate drafts for the large model.
However, since the small model still generates through inefficient autoregressive decoding, this process could limit the efficiency of speculative decoding due to the drafting latency~\cite{chen2024cascade}.
As mentioned in Section~\ref{sec:Motivation}, the small model's output often overlaps substantially with the existing context or retrieved code snippets.
Inspired by this, instead of relying on another smaller neural model to generate the draft for the small model, we apply context-based matching to obtain a draft, which incurs negligible latency and computational resource costs compared to neural language models.
\rev{
Table~\ref{tab:context_based_matching} presents a concrete example to help illustrate the context-based matching.
When the user is writing a new line, we identify the nearest non-empty line before the completion point (e.g. \textit{scheduler = ...} in the example) and match it against both the context and the retrieved code snippets. 
If a match is found, we extract its next non-empty, non-comment statement (e.g. \textit{classifier = ...} in the example) as the draft.
}
For in-line completion tasks where the user has partially written a line, we instead use the current partial line to find lines that start with the same fragment, and if a match is found, we take the remaining suffix of that statement as the draft.

If the completion from the small model is rejected, its suggestion is then used to initiate speculative decoding in the large model, effectively creating a two-stage speculative decoding process. 
This dual-stage acceleration mechanism allows both models to benefit from faster decoding, resulting in a low-latency pipeline.

\subsection{Iterative Retrieval}
Code repositories often contain multiple files with similar patterns or relevant structures.
To exploit this, we follow prior work~\cite{wangRLCoderReinforcementLearning2024} to build a retrieval corpus $RC_{repo}$ by splitting other files in the repository into code snippets using blank lines as delimiters.

At the initial retrieval, we form a query $q$ by concatenating the last lines of the left context with the first lines of the right context, as empirical evidence suggests that combining both contexts leads to better retrieval performance.
\rev{
We then employ BM25 to score candidate snippets with respect to the query and select the top-$K$ ones to prepend to the model input.
BM25 is chosen for two key reasons: (1) it is more efficient than dense retrieval methods, and (2) prior work~\cite{wuRepoformerSelectiveRetrieval2024, zhang2023repocoder} has shown that sparse retrievers are competitive with dense retrievers in code completion.
}

To further enrich the context, we introduce an iterative retrieval mechanism.
This component is motivated by two key observations. 
First, a more informative context often leads to better answer quality~\cite{rezaei2025vendi}. 
Second, as shown in Section~\ref{sec:Motivation}, even when the user rejects the small model’s suggestion, it can still provide valuable information about the developer's intent. 
To fully utilize the small model's output, we propose using it to perform a second round of retrieval, thereby adapting the context with new information.

To achieve iterative retrieval, a direct solution is to append the small model's output $\hat{Y}_{\mathcal{SM}}$ to the original query $q$ for retrieval.
However, $\hat{Y}_{\mathcal{SM}}$ is typically much shorter than $q$, its contribution to the overall query is limited, resulting in only marginal improvements.
To address this limitation, we design a weighted scoring mechanism that explicitly amplifies the impact of the small model's output.
Specifically, we compute similarity scores using the original query $q$ and the small model's output $\hat{Y}_{\mathcal{SM}}$ separately, denoted as $Score_{context}$ and $Score_{\mathcal{SM}}$.
Yet, naively summing these two scores can introduce negative effects when the small model produces misleading outputs.
To mitigate this issue, our weighting mechanism dynamically adjusts the contribution of $\hat{Y}_{\mathcal{SM}}$ according to its confidence $w_{conf}$, ensuring that low-confidence outputs exert less influence on the final score.
After normalizing each score vector, the combined score is computed as:
\begin{equation}
    Score_{adaptive}^i = \frac{Score_{context}^i}{\sum_j{Score_{context}^j}} + w_{conf} \cdot \frac{Score_{\mathcal{SM}}^i}{\sum_j{Score_{\mathcal{SM}}^j}}
\end{equation}
Finally, the snippets are retrieved according to $Score_{adaptive}$, and only the top-$k$ snippets are incorporated into the LLM input. 
This procedure allows the model to primarily incorporate the adaptive information suggested by the small model, while also leveraging the stable context from the original retrieval, thereby yielding richer and more relevant context for code generation.

\section{experimental Setup}\label{sec:ExperimentalSetup}
This section describes the datasets, the evaluation metrics used to evaluate our approach, and the implementation details of our approach.

\subsection{Small model training details}
To implement and evaluate \appname, we need a lightweight model that can run entirely on user devices and perform well in code completion.
Because the small model's effectiveness significantly affects the effectiveness of model cascading.
Specifically, if its performance is too weak, most requests will fall back to the large model, resulting in minimal latency reduction.
Unfortunately, we cannot find any publicly available small model that meets the requirements.
Therefore, we train a small model specifically optimized for the line-level code completion task.

\subsubsection{Training set}~\label{sec:training_set}
We used the Python subset of The Stack V2~\cite{lozhkov2024starcoder} as our training corpus.
To enhance code quality and minimize noise, we perform several preprocessing steps prior to training, as done in previous work~\cite{guo2024deepseek}.
After applying these preprocessing steps, we obtained a total of 41M samples for training and 41K samples for validation.
These preprocessing steps help ensure that the training data is both syntactically valid and semantically useful, thereby improving model quality and efficiency.

\subsubsection{Small model construction}
We build our small model upon the LLaMA architecture, following the configurations proposed by Hu et al.~\cite{huMiniCPMUnveilingPotential2024}, with the only modification being an adjustment of the vocabulary size to 16,256.
To identify an optimal balance between model size and performance, we conduct preliminary experiments using four model scales: 70M, 0.1B, 0.17B, and 0.2B parameters. 
Among them, the 0.1B model demonstrates the best trade-off between computational efficiency and accuracy.

\subsubsection{Small model training}
To obtain a more compact vocabulary, we train a custom tokenizer on our preprocessed training corpus.
For the training objective, we use the standard cross-entropy loss over the model's token predictions.
To equip the model with both next-token prediction (NTP) and infilling capabilities, we follow the training strategy proposed by Bavarian et al.~\cite{bavarianEfficientTrainingLanguage2022}, using 10\% of the training data in NTP format and 90\% in fill-in-the-middle (FIM) format.
Specifically, we adopt only the SPM (suffix-prefix-middle) mode for FIM, as we find it more effective for code completion tasks.
We pretrain the model on the full training set, followed by fine-tuning on a curated subset. 
This subset comprises 75\% of the original pretraining data and 25\% synthetic code samples generated by OpenCoder~\cite{huangOpenCoderOpenCookbook2024}, a ratio chosen to maintain consistency in data distribution while introducing diversity.
Additionally, in 10\% of the training samples, the suffix is removed during FIM formatting to encourage the model to generate completions without relying on future context, a proportion we found sufficient for effective learning without performance degradation.
For optimization, we use AdamW with an initial learning rate of $3 \times 10^{-4}$, cosine learning rate decay with a minimum learning rate, and 1000 warmup steps.
Through this training procedure, we obtain our final small model optimized for practical code completion scenarios.

\subsection{Evaluation Setup}
\subsubsection{Baselines}
To evaluate the effectiveness and efficiency of \appname, we compare it against four representative baselines, each reflecting a distinct design in code completion systems.
\begin{itemize}
    \item \textbf{SLM-only} uses only the small local model for code completion. 
    It represents the most cost- and latency-efficient approach, at the expense of accuracy.
    \item \textbf{LLM-only} uses the large model directly to generate completions. 
    It reflects the design of existing high-performance systems such as GitHub Copilot and provides strong accuracy at the cost of significantly higher inference latency and resource usage.
    \item \textbf{$\text{SLM}_\text{twice}$} simulates a fallback strategy where the small model is invoked a second time if the initial suggestion is rejected. 
    To approximate this in our offline evaluation, we use beam search with a size of 2 and select the better candidate based on its similarity to the ground truth.
    This baseline is designed to assess whether invoking the small model again is a viable alternative to switching to the large model, thus helping justify the need for model cascading.
    \item  \textbf{$\text{LLM}_\text{twice}$} simulates a strategy where the large model is called twice. 
    Like the small-model variant, we approximate this with beam search and oracle selection.
    This design provides a reference point to examine the efficiency-accuracy tradeoff achieved by model cascading.
\end{itemize}

We conduct the experiments with three state-of-the-art 7B Code LLMs, i.e., \textit{Qwen2.5-Coder-7B}~\cite{hui2024qwen2}, \textit{DeepSeek-Coder-7B-base}~\cite{guo2024deepseek}, and \textit{CodeLLama-7B}~\cite{roziere2023code}.
For convenience, we refer to the three LLMs as \textit{QC}, \textit{DSC}, and \textit{CL}, respectively.
We use the base versions of these LLMs following prior work~\cite{wangRLCoderReinforcementLearning2024}.
We chose 7B-scale models for two reasons.
First, according to the feedback from industry practitioners, current code completion systems deployed in real-world development environments commonly rely on 7B models due to their balance of performance and cost.
Second, recent findings~\cite{liu2024graphcoder} suggest that in repository-level code completions, performance does not scale significantly with model size because of limited intra-repository knowledge, and the performance gain from 7B to 16B is marginal.

Moreover, we also compare MCCom with RepoCoder~\cite{zhang2023repocoder} and CSDrafting~\cite{chen2024cascade} since they are highly relevant to our work.
\begin{itemize}
    \item \textbf{RepoCoder} is a retrieval-augmented code completion method, where code retrieval and model generation are performed iteratively.
    This resembles our use of iterative retrieval; however, in contrast to RepoCoder, our framework uses a lightweight local model to generate the first-round completion, rather than relying exclusively on a large model.
    \item \textbf{CSDrafting} explores speculative decoding in a cascaded multi-model setting, where the lowest-level generator is a bigram language model augmented with prompt lookup. 
    This approach is conceptually related to our two stage speculative decoding. 
    However, unlike our framework, where the local model provides fast responses directly to users, CSDrafting deploys all models on the same device, with the smaller model serving solely to draft outputs for the larger one.
\end{itemize}
For CSDrafting, we only report results on Qwen and DeepSeek models, as CodeLlama lacks sufficiently small variants (e.g., below 3B parameters) to enable cascaded speculative decoding.
Specifically, we adopt Qwen2.5-Coder-0.5B as the draft model for Qwen2.5-Coder-7B, and DeepSeek-Coder-1.3B as the draft model for DeepSeek-Coder-6.7B.

\begin{table}[t]
    \centering
    \caption{Statistic of Datasets}
    \label{tab:datasets}
    \begin{threeparttable}
    \begin{tabular}{ccccc}
        \toprule
        \textbf{Benchmark}&\textbf{Category}&\textbf{\#Samples}&\textbf{Avg.Lines}&\textbf{Avg.Tokens}\\
        \multirow{2}{*}{RepoEval}&Python(Line)&1600&1.00&15.03\\
        &Python(API)&1600&2.48&34.91\\
        \midrule
        \multirow{2}{*}{StmtEval}&Python(Line)&999&2.17&17.69\\
        &Python(suffix)&999&2.09&9.93\\
        \bottomrule
    \end{tabular}
    \begin{tablenotes}
        \footnotesize
        \item[1] \textit{\#Samples} denotes the number of samples in each benchmark. 
        \textit{Avg.Lines} and \textit{Avg.Tokens} indicate the average number of lines and tokens in the target code snippets, respectively. Tokenization is performed using the tokenizer of DeepSeekCoder-1.3B.
    \end{tablenotes}
    \end{threeparttable}
\end{table}

\subsubsection{Evaluation Datasets}\label{subsec:benchmark}
As the first to explore model cascading between local and cloud models for code completion, we focus on the line-level code completion task for two main reasons. 
\textit{First}, \appname aims to reduce latency, and line-level completion is better suited for this goal since it is inherently more interactive and places stricter requirements on latency. 
\textit{Second}, line-level represents an important and prevalent scenario in real-world deployments. 
A survey of 599 professionals across 18 IT companies~\cite{wangPractitionersExpectationsCode2023} indicated significantly higher adoption for line-level (45\%) than for block-level (16\%). 
Furthermore, internal telemetry from Tencent reveals that line-level tasks account for 80\% of requests during software maintenance. 
This is a critical stage that consumes 40–80\% of development budgets~\cite{glass2002facts}.

We use the RepoEval benchmark for evaluation, which is widely used to evaluate code completion approaches~\cite{wangRLCoderReinforcementLearning2024, wuRepoformerSelectiveRetrieval2024}.
RepoEval~\cite{zhang2023repocoder} consists of code completion instances from 32 Python repositories.
As this work focuses on line-level code completion, we use its line-level and API-level subsets for evaluation.
However, each sample in RepoEval corresponds to a code line, which is not always a functionally complete statement.
As a result, some targets offer limited practical utility to developers.
For example, we found six samples in RepoEval where the target is simply an opening brace preceded by indentation (e.g., \textit{"\textvisiblespace\textvisiblespace\textvisiblespace\textvisiblespace\{"}).
Such cases assess a model’s ability to reproduce syntactically incomplete lines rather than generate meaningful statements, thereby introducing potential evaluation bias.

To complement RepoEval, we construct a new benchmark, StmtEval, which considers a full statement as a "line".
We begin by carefully curating a collection of Python repositories from GitHub that meet the following criteria: they were created after January 1, 2024\footnote{The Stack v2 only collected data up to September 14, 2023.}, are non-forked, and have more than 100 stars.
To ensure diversity, we balance the number of machine learning and non-machine learning projects by identifying commonly used packages such as \texttt{pytorch}, \texttt{tensorflow}, and others in the \texttt{import} statements.
From each selected repository, we randomly sample up to 15 statements, excluding lines that are code comments. 
In total, we obtain 999 statements drawn from 93 repositories for code completion.
To simulate realistic usage scenarios, we further split each statement at a random character position to reflect situations where users may request code completions at arbitrary points. 
We refer to the complete statements as \textit{StmtEval-full}, and the randomly truncated statements as \textit{StmtEval-suffix}.
To better evaluate model performance under different practical settings, each benchmark is tested under two configurations: one where both left and right contexts are provided, and another where only the left context is available.

Table~\ref{tab:datasets} presents the statistics of the two benchmarks. 
For convenience, we use \textit{RE} and \textit{SE} as the abbreviations of RepoEval and StmtEval, respectively.

\subsubsection{Evaluation Metrics}~\label{sec:metric}
We use two widely adopted metrics in code completion research, i.e., \textit{\textbf{Exact Match (EM)}} and \textit{\textbf{Edit Similarity (ES)}}~\cite{levenshtein1966binary}, consistent with prior studies~\cite{ding2024cocomic, zhang2023repocoder, wangRLCoderReinforcementLearning2024, eghbali2024hallucinator}.
EM quantifies prediction accuracy by determining whether the generated code exactly matches the ground truth. 
ES is defined as:
\begin{equation}
ES(\hat{Y}, Y) = \frac{1 - \text{Lev}(\hat{Y}, Y)}{\max(|\hat{Y}|, |Y|)},
\end{equation}
where \textit{Lev} denotes the Levenshtein distance~\cite{levenshtein1966binary}, and $|\cdot|$ represents sequence length. 
This metric measures the normalized similarity between the generated output $\hat{Y}$ and the reference $Y$, reflecting how many edits are required to transform one into the other. 
For readability, we report EM and ES scores multiplied by 100 in all result tables.

\subsection{Implementation Details}
In our experiments, the large model $\mathcal{LM}$ is hosted on an NVIDIA A800 GPU with 80GB of memory, while the small model $\mathcal{SM}$ is deployed on a local NVIDIA RTX 3080Ti with 12GB of memory. 
Both models are configured with a maximum input context length of 2048 tokens, with up to 512 tokens allocated for retrieval content, and a generation limit of 128 tokens.
Generation is performed using greedy decoding, and sampling-based methods are disabled to ensure deterministic outputs. 
During inference, we also apply task-specific stopping criteria to avoid unnecessary computation. 
On the RepoEval-Line benchmark, decoding stops once a complete line is generated. 
For all other benchmarks, generation terminates upon completing a syntactically valid statement.
In our routing strategy, the local model computes a confidence score based on the first $N=3$ generated tokens and the threshold is set to 0.8.
To simulate whether a user would accept a given completion, we adopt a simple exact match strategy: if the generated completion exactly matches the ground truth, it is considered accepted.
Otherwise, it is treated as rejected.

\section{Evaluation Results}\label{sec:EvalutaionResults}
In this section, we report and analyze the experimental results to answer the following research questions (RQs):
\begin{itemize}
    \item RQ1: How efficient is \appname\ compared to the baselines?
    \item RQ2: How effective is \appname\ in line-level code completion?
    \item RQ3: Does each component of \appname\ contribute to its performance?
\end{itemize}

\subsection{RQ1: Efficiency of \appname}\label{subsec:RQ1}
\begin{table*}[t]
    \centering
    \caption{Comparison of efficiency (in milliseconds)}
    \label{tab:efficiency}
    \begin{tabular}{lllllllll}
        \toprule
        & \multicolumn{4}{c}{\textbf{Left \& Right Contexts}} & \multicolumn{4}{c}{\textbf{Only Left Context}} \\
        \textbf{Method} & \textbf{\makecell{RE\\(Line)}} & \textbf{\makecell{RE\\(API)}} & \textbf{\makecell{SE\\(Line)}} & \textbf{\makecell{SE\\(Suffix)}} &
         \textbf{\makecell{RE\\(Line)}} & \textbf{\makecell{RE\\(API)}} & \textbf{\makecell{SE\\(Line)}} & \textbf{\makecell{SE\\(Suffix)}} \\
        \midrule
        SLM & 235 & 423 & 295 & 186 & 247 & 465 & 295 & 197 \\
        SLM$_\text{twice}$ & 330 & 731 & 428 & 306 & 327 & 784 & 438 & 298 \\
        \midrule
        CL & 675 & 1015 & 732 & 586 & 643 & 1045 & 833 & 687 \\
        CL$_\text{twice}$ & 1008 & 1777 & 1173 & 949 & 941 & 1765 & 1240 & 1001 \\
        CL$_\text{repocoder}$ & 1178 & 1886 & 1320 & 997 & 1122 & 1965 & 1509 & 1202 \\
        CL$_\text{\appname}$ & \textbf{433} & \textbf{815} & \textbf{562} & \textbf{314} & \textbf{516} & \textbf{920} & \textbf{730} & \textbf{418} \\
        \midrule
        DSC & 703 & 1072 & 770 & 599 & 685 & 1179 & 852 & 675 \\
        DSC$_\text{twice}$ & 1031 & 1758 & 1157 & 927 & 956 & 1751 & 1114 & 892 \\
        DSC$_\text{repocoder}$ & 1210 & 1960 & 1351 & 1002 & 1174 & 2142 & 1477 & 1128 \\
        DSC$_\text{CSDrafting}$ & 781 & 1012 & 878 & 822 & 751 & 934 & 793 & 703 \\
        DSC$_\text{\appname}$ & \textbf{435} & \textbf{812} & \textbf{545} & \textbf{312} & \textbf{529} & \textbf{902} & \textbf{675} & \textbf{399} \\
        \midrule
        QC & 588 & 811 & 592 & 501 & 569 & 831 & 646 & 527 \\
        QC$_\text{twice}$ & 883 & 1920 & 1408 & 1187 & 783 & 1511 & 1075 & 882 \\
        QC$_\text{repocoder}$ & 979 & 1430 & 989 & 800 & 938 & 1466 & 1081 & 848 \\
        QC$_\text{CSDrafting}$ & 830 & 933 & 810 & 745 & 807 & 1028 & 861 & 748 \\
        QC$_\text{\appname}$ & \textbf{408} & \textbf{719} & \textbf{495} & \textbf{293} & \textbf{483} & \textbf{783} & \textbf{607} & \textbf{358} \\
        \bottomrule
    \end{tabular}
\end{table*}

\begin{figure*}[t]
    \centering
    \includegraphics[width=0.9\linewidth]{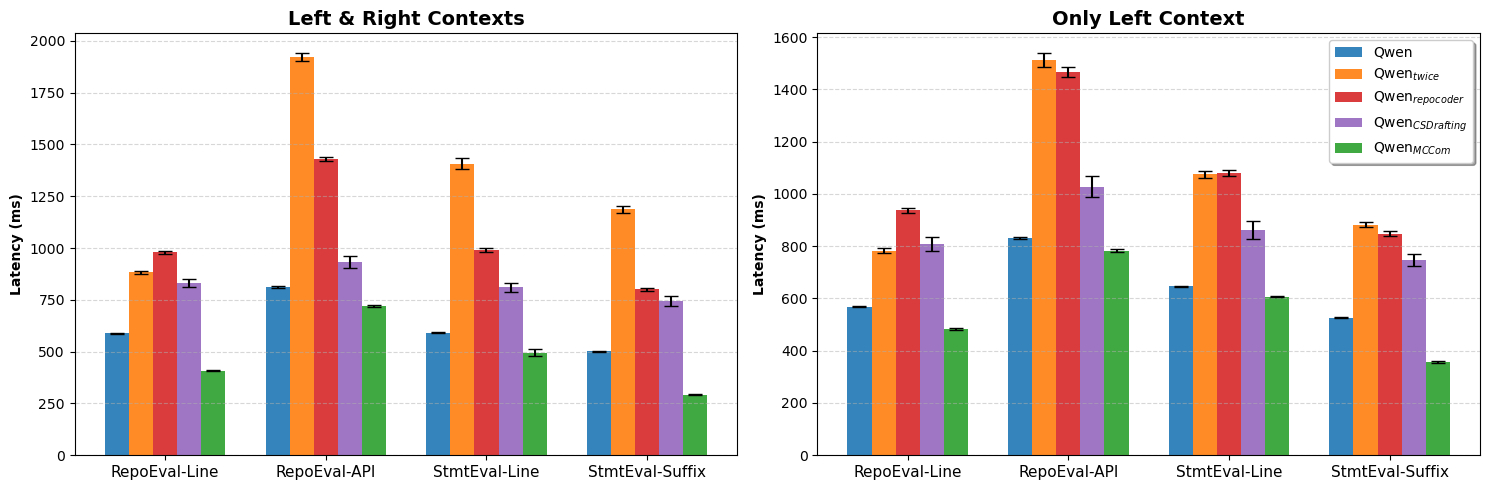}
    \caption{Statistical Information about Efficient Results}
    \label{fig:statistical}
\end{figure*}
To validate whether \appname\ can balance the latency and the accuracy of code completion, we first evaluate and compare the efficiency of \appname\ and the baselines.
To reflect realistic usage, we simulate the end-to-end workflow of modern code completion tools, which typically includes local context gathering, network transmission, and cloud-based model inference~\cite{sombanerjeeGitHubCopilotChat2025}. 
Accordingly, our evaluation considers retrieval, a query latency of 200 ms when invoking the cloud model, and model inference. 
For the LLM$_\text{twice}$ baseline, this network latency is counted only once, as both completions are assumed to be generated within a single request using beam search.
\rev{
To ensure the reliability of our results, we repeated each experiment five times and reported the average latency in Table~\ref{tab:efficiency}. 
Furthermore, we computed the 95\% confidence intervals using the t-distribution, confirming that the performance gaps between \appname\ and baselines are statistically significant.
Due to space constraints, we present the visualizations for Qwen2.5-Coder-7B as a representative example in Figure~\ref{fig:statistical}, and the complete visualizations is available in our repository (refer to Section~\ref{sec:DataAvailability}).
}

\textbf{Compared to LLM-only and LLM$_\text{twice}$}, \appname consistently achieves lower latency across all test scenarios.
Specifically, \appname achieves 5.8\% to 47.9\% lower latency than LLM-only, and shows more substantial gains over LLM$_\text{twice}$, with speed improvements ranging from 38.3\% to 75.3\%.
For example, on StmtEval-Suffix (left\&right context) with QC as the large model, \appname\ achieves a latency of 293 ms, reducing the delay by 41.4\% compared to LLM-only.
This result highlights the effectiveness of model cascading in reducing latency by leveraging the small model whenever possible.

\appnamebold\textbf{ also outperforms both RepoCoder and CSDrafting}, reducing average latency by 57.1\% compared to RepoCoder and by 35.9\% compared to CSDrafting.
RepoCoder is significantly slower because it invokes the cloud model twice, whereas \appname\ requires at most a single cloud call. 
Surprisingly, CSDrafting is not only slower than \appname\ but also slower than directly using the large model, despite employing speculative sampling. 
Two main factors account for this inefficiency: (1) the outputs from its lowest-level draft generator have a relatively low acceptance rate (14.3\%), which constrains their usefulness; and (2) since CSDrafting only allows the target model to review the draft, which is ill-suited for code completion scenarios where the input is usually much longer than the output, making it difficult to amortize the overhead introduced by the draft model.

In contrast, our framework avoids these issues. 
Unlike CSDrafting, \appname\ generates a high-quality draft by exactly matching statements from code snippets and achieves an acceptance rate of 51.5\%.
Moreover, the primary role of the local model is to provide users with fast responses, while using its outputs as a draft for the cloud model is merely a byproduct.
Once the cloud model validates this draft, it immediately continues generation on its own, without waiting for the local model to generate additional draft tokens. 
As a result, \appname\ achieves low latency without suffering from the limitations observed in CSDrafting.

We also observe that latency gains are less pronounced under left-context-only settings, especially for RepoEval-API.
When both left and right contexts are provided, \appname\ achieves significant latency improvements compared to LLM-only, ranging from 11.4\% to 47.9\%.
In contrast, with only the left context, improvements are more modest, ranging from 5.8\% to 40.9\%.
This is primarily because the small model has a lower accuracy in such scenarios, limiting the reduction in large model invocations.
\appname\ delivers less improvement on RepoEval-API because small models are not well-suited to handling the diverse range of API calls, which make up the RepoEval-API dataset.
All of these API calls are in the repository and require understanding project-specific context to infer which API to call and how to configure its parameters, which exceeds the capacity of the small model.
In practice, user requests are more diverse, with such a scenario, where there are only API call completions, is unlikely to occur.
Therefore, we expect the actual speedup in practice would be more significant than that in RepoEval-API.

\noindent \fbox{
	\parbox{0.95\linewidth}{\textbf{RQ1 Summary:} \appname is significantly faster than both LLM-only and LLM$_\text{twice}$ baselines, achieving an average latency reduction of 25.6\% and 53.6\% respectively, validating its efficiency.
    Moreover, \appname\ outperforms RepoCoder and CSDrafting by effectively leveraging the characteristics of code completion tasks and the local–cloud cascading scenario.
    }
}

\subsection{RQ2: Effectiveness of \appname}\label{subsec:RQ2}
\begin{table}[t]
    \centering
    \caption{Comparison of effectiveness using left \& right context.}
    \label{tab:baseline_prefix_and_suffix}
    \begin{threeparttable}
    \begin{tabular}{lccccccccccc}
        \toprule
        & \multicolumn{2}{c}{\textbf{RE (Line)}} & \multicolumn{2}{c}{\textbf{RE (API)}} & \multicolumn{2}{c}{\textbf{SE (Line)}} & \multicolumn{2}{c}{\textbf{SE (Suffix)}} & \multicolumn{3}{c}{\textbf{Average}}  \\
        \textbf{Method} & \textbf{EM} & \textbf{ES} & \textbf{EM} & \textbf{ES} & \textbf{EM} & \textbf{ES} & \textbf{EM} & \textbf{ES} & \textbf{EM} & \textbf{ES} & \textbf{time} \\
        \midrule
        SLM & 54.8 & 80.8 & 40.9 & 75.4 & 53.0 & 80.2 & 72.1 & 89.4 & 55.2 & 81.4 & 285 \\
        SLM$_\text{twice}$ & 61.8 & 84.9 & 49.7 & 80.7 & 60.2 & 84.4 & 77.8 & 91.9 & 62.3 & 85.5 & 449 \\
        \midrule
        CL & 61.9 & 82.0 & 53.6 & 77.8 & 61.3 & 79.5 & 77.8 & 90.9 & 63.6 & 82.6 & \underline{752} \\
        CL$_\text{twice}$ & \textbf{68.9} & \textbf{86.8} & \textbf{59.3} & \textbf{82.8} & \underline{67.3} & \textbf{84.7} & \textbf{81.1} & \textbf{92.4} & \textbf{69.2} & \textbf{86.7} & 1227 \\
        CL$_\text{repocoder}$ & 62.1 & 82.0 & 54.1 & 77.9 & 61.2 & 79.7 & 78.0 & 91.1 & 63.8 & 82.7 & 1345 \\
        CL$_\text{MCCom}$ & \underline{68.7} & \underline{85.4} & \underline{58.9} & \underline{81.1} & \textbf{67.6} & \underline{82.7} & \textbf{81.1} & \textbf{92.4} & \underline{69.1} & \underline{85.4} & \textbf{531} \\
        \midrule
        DSC & 62.8 & 81.5 & 54.6 & 78.9 & 62.7 & 79.3 & 80.3 & 92.6 & 65.1 & 83.1 & \underline{786} \\
        DSC$_\text{twice}$ & \underline{70.1} & \textbf{87.7} & \textbf{62.3} & \textbf{84.4} & \underline{70.4} & \textbf{85.9} & \textbf{83.4} & \textbf{93.7} & \textbf{71.5} & \textbf{87.9} & 1218 \\
        DSC$_\text{repocoder}$ & 63.1 & 81.5 & 55.3 & 79.1 & 62.7 & 79.1 & 80.0 & 92.4 & 65.3 & 83.0 & 1381 \\
        DSC$_\text{MCCom}$ & \textbf{70.3} & \underline{85.3} & \underline{60.5} & \underline{82.2} & \textbf{70.6} & \underline{85.0} & \underline{82.6} & \underline{93.5} & \underline{71.0} & \underline{86.5} & \textbf{526} \\
        \midrule
        QC & 66.5 & 85.0 & 60.1 & 81.7 & 65.8 & 83.3 & 79.1 & 92.3 & 67.9 & 85.6 & \underline{623} \\
        QC$_\text{twice}$ & \textbf{72.2} & \textbf{88.7} & \textbf{65.4} & \textbf{85.6} & \underline{70.8} & \textbf{86.9} & \underline{81.5} & \underline{93.0} & \underline{72.5} & \textbf{88.5} & 1349 \\
        QC$_\text{repocoder}$ & 66.2 & 84.6 & 60.4 & 81.9 & 66.3 & 83.5 & 79.6 & 92.5 & 68.1 & 85.6 & 1050 \\
        QC$_\text{MCCom}$ & \underline{71.6} & \underline{87.3} & \underline{65.3} & \underline{84.8} & \textbf{71.2} & \underline{86.3} & \textbf{82.5} & \textbf{93.5} & \textbf{72.6} & \underline{88.0} & \textbf{479} \\
        \bottomrule
    \end{tabular}
    \begin{tablenotes}
     \small
        \item *The best results are highlighted in bold. The second-best results are indicated with underlining.
    \end{tablenotes}
    \end{threeparttable}
\end{table}

\begin{table}[t]
    \centering
    \caption{Comparison of effectiveness using only the left context.}
    \label{tab:baseline_prefix}
    \begin{threeparttable}
    \begin{tabular}{lccccccccccc}
        \toprule
        & \multicolumn{2}{c}{\textbf{RE (Line)}} & \multicolumn{2}{c}{\textbf{RE (API)}} & \multicolumn{2}{c}{\textbf{SE (Line)}} & \multicolumn{2}{c}{\textbf{SE (Suffix)}} & \multicolumn{3}{c}{\textbf{Average}} \\
        \textbf{Method} & \textbf{EM} & \textbf{ES} & \textbf{EM} & \textbf{ES} & \textbf{EM} & \textbf{ES} & \textbf{EM} & \textbf{ES} & \textbf{EM} & \textbf{ES} & \textbf{time} \\
        \midrule
        SLM & 34.6 & 63.9 & 26.8 & 61.4 & 28.3 & 59.8 & 60.8 & 82.3 & 37.6 & 66.8 & 301 \\
        SLM$_\text{twice}$ & 39.8 & 69.1 & 32.3 & 67.2 & 33.9 & 65.1 & 66.6 & 85.7 & 43.1 & 71.8 & 462 \\
        \midrule
        CL & 45.2 & 70.9 & 38.1 & 66.2 & 38.7 & 63.5 & 67.1 & 83.7 & 47.3 & 71.1 & \underline{802} \\
        CL$_\text{twice}$ & \textbf{51.7} & \textbf{76.4} & \textbf{43.9} & \textbf{72.6} & \textbf{45.7} & \textbf{70.4} & \underline{70.9} & \textbf{87.0} & \textbf{53.0} & \textbf{76.6} & 1237 \\
        CL$_\text{repocoder}$ & 45.8 & 71.0 & 40.0 & 67.4 & 39.4 & 63.5 & 67.7 & 84.1 & 48.2 & 71.5 & 1450 \\
        CL$_\text{MCCom}$ & \underline{50.7} & \underline{74.1} & \underline{43.3} & \underline{69.2} & \underline{43.8} & \underline{66.6} & \textbf{72.2} & \underline{86.4} & \underline{52.5} & \underline{74.1} & \textbf{646} \\
        \midrule
        DSC & 47.4 & 72.9 & 41.5 & 71.5 & 43.8 & 69.0 & 69.3 & 86.2 & 50.5 & 74.9 & \underline{848} \\
        DSC$_\text{twice}$ & \textbf{54.3} & \textbf{77.3} & \underline{45.8} & \textbf{75.2} & \textbf{49.9} & \textbf{73.3} & \textbf{74.3} & \textbf{89.0} & \textbf{56.1} & \textbf{78.7} & 1178 \\
        DSC$_\text{repocoder}$ & 48.5 & 73.5 & 43.4 & 72.4 & 44.6 & 69.1 & 69.6 & 86.5 & 51.5 & 75.4 & 1481 \\
        DSC$_\text{MCCom}$ & \underline{52.4} & \underline{75.6} & \textbf{46.3} & \underline{74.3} & \underline{47.7} & \underline{71.1} & \underline{73.3} & \underline{88.1} & \underline{54.9} & \underline{77.2} & \textbf{626} \\
        \midrule
        QC & 51.8 & 75.8 & 48.3 & 75.7 & 47.7 & 72.2 & 69.9 & 87.3 & 54.4 & 77.7 & \underline{643} \\
        QC$_\text{twice}$ & \textbf{57.5} & \textbf{79.0} & \textbf{53.1} & \textbf{78.4} & \textbf{51.5} & \textbf{74.9} & \underline{73.9} & \textbf{89.0} & \textbf{59.0} & \textbf{80.3} & 1063 \\
        QC$_\text{repocoder}$ & 53.0 & 76.4 & 49.2 & 76.4 & 48.3 & 72.4 & 70.7 & 87.4 & 55.3 & 78.1 & 1083 \\
        QC$_\text{MCCom}$ & \underline{56.1} & \underline{78.0} & \underline{51.5} & \underline{77.2} & \underline{50.7} & \underline{73.9} & \textbf{74.2} & \textbf{89.0} & \underline{58.1} & \underline{79.5} & \textbf{558} \\
        \bottomrule
    \end{tabular}
    \begin{tablenotes}
     \small
        \item *The best results are highlighted in bold. The second-best results are indicated with underlining.
    \end{tablenotes}
    \end{threeparttable}
\end{table}

In this RQ, we aim to investigate whether \appname can maintain the accuracy of code completion.
Specifically, we evaluate and compare \appname and the baselines in terms of EM and ES.
It is worth noting that CSDrafting does not alter the model outputs, and therefore, they have the same performance as Qwen.
As shown in Table~\ref{tab:baseline_prefix_and_suffix} and Table~\ref{tab:baseline_prefix}, \appname\ consistently outperforms SLM-only, SLM$_\text{twice}$, LLM-only, and RepoCoder across all datasets and LLM. 

\textbf{Compared to SLM$_\text{twice}$}, \appname delivers substantial performance gains with only a modest increase in latency. 
On average, it achieves a 29.1\% improvement in exact match accuracy while incurring merely a 17.5\% increase in latency, demonstrating a favorable accuracy–efficiency trade-off. 

\textbf{Compared to LLM-only}, \appname\ achieves an EM improvement of 2.9\%–13.5\% while substantially reducing inference latency, striking a well-balanced trade-off between efficiency and accuracy. 
These accuracy gains can be attributed to two factors. 
First, there exist cases where the large model fails but the small model succeeds; since \appname\ employs the SLM first and incorporates implicit human feedback, it can successfully solve such cases. 
Second, the interactive retrieval mechanism in \appname\ enables the large model to handle more cases correctly, as we will further demonstrate in Section~\ref{sec:rq3}.  

\textbf{Compared to LLM$_\text{twice}$}, \appname\ may not achieve higher accuracy in most settings; however, this is acceptable given the substantial computational overhead of LLM$_\text{twice}$, which makes it impractical for real-world deployment.
On average, \appname\ reduces latency by 53.6\% while incurring only 1.0\% decrease in EM compared to LLM$_\text{twice}$, highlighting its favorable accuracy–efficiency trade-off.
Notably, in certain scenarios \appname\ even outperforms this strongest baseline. 
For instance, with deepseekcoder as the backbone model on the RepoEval API benchmark under the left-only context setting, \appname\ achieves a 0.6 percentage point improvement in exact match accuracy over LLM$_\text{twice}$. 
This phenomenon may stem from the complementary knowledge of the small and large models, which in some cases leads to better results than directly applying beam search with the large model alone. 

\textbf{Compared to RepoCoder}, \appname\ achieves an EM improvement of 3.3\%–12.6\% while simultaneously reducing latency. 
The key distinction lies in how intermediate generations are treated: RepoCoder performs iterative retrieval and generation in a fully automatic manner, which can overwrite a correct early prediction with an incorrect later one due to the absence of human interaction. 
For example, on the RepoEval-API subset with left and right contexts and Qwen2.5-Coder-7B, we observed that 1.7\% of samples changed from incorrect to correct through later iterations, but 1.4\% of samples degraded from correct to incorrect, resulting in little net improvement in overall performance.
In contrast, \appname\ presents intermediate completions from the local model directly to the user and leverages implicit human feedback to decide whether further retrieval and generation are necessary. 
This interactive process not only reduces redundant cloud calls but also mitigates the risk of regression from over-iteration, thereby yielding a more favorable balance between accuracy and efficiency.

\noindent \fbox{
	\parbox{0.95\linewidth}{\textbf{RQ2 Summary:} On average, \appname improves accuracy by 8.9\% over LLM-only across benchmarks and by 7.6\% over RepoCoder.
    Furthermore, it strikes a better balance between latency and accuracy than SLM$_\text{twice}$ and LLM$_\text{twice}$, validating the rationale behind our design of combining the small and large models.}
}

\subsection{RQ3: Contributions of Each Component}\label{sec:rq3}
\begin{table}[t]
    \centering
    \caption{Ablation Study Results on RepoEval and StmtEval}
    \label{tab:ablation}
    \begin{threeparttable}
    \begin{tabular}{lccccccccccc}
        \toprule
        & \multicolumn{2}{c}{\textbf{RE (Line)}} & \multicolumn{2}{c}{\textbf{RE (API)}} & \multicolumn{2}{c}{\textbf{SE (Line)}} & \multicolumn{2}{c}{\textbf{SE (Suffix)}} & \multicolumn{3}{c}{\textbf{Average}} \\
        \textbf{Method} & \textbf{EM} & \textbf{time} & \textbf{EM} & \textbf{time} & \textbf{EM} & \textbf{time} & \textbf{EM} & \textbf{time} & \textbf{EM} & \textbf{time} & \textbf{Eff.} \\
        \midrule
        \rowcolor{gray!50}
        \multicolumn{12}{c}{Left \& Right Context} \\
        \midrule
        SLM & 54.8 & 235 & 40.9 & 423 & 53.0 & 295 & 72.1 & 186 & 55.2 & 285 & -- \\
        \quad w 1st SD & 54.8 & 175 & 40.9 & 292 & 53.0 & 241 & 72.1 & 153 & 55.2 & 215 & -- \\
        \hdashline
        QC$_\text{\appname}$ & 71.6 & 408 & 65.3 & 719 & 71.2 & 495 & 82.5 & 293 & 72.6 & 479 & 40.8 \\
        \quad wo IR & 71.1 & 407 & 64.3 & 721 & 70.3 & 488 & 82.0 & 293 & 71.9 & 477 & 39.3 \\
        \quad wo 1st SD & 71.6 & 482 & 65.3 & 875 & 71.2 & 565 & 82.5 & 343 & 72.6 & 566 & 28.4 \\
        \quad wo 2nd SD & 71.6 & 440 & 65.3 & 802 & 71.2 & 529 & 82.5 & 311 & 72.6 & 520 & 33.8 \\
        \quad wo SD & 71.6 & 516 & 65.3 & 959 & 71.2 & 600 & 82.5 & 361 & 72.6 & 609 & 24.7 \\
        \midrule
        \rowcolor{gray!50}
        \multicolumn{12}{c}{Only Left Context} \\
        \midrule
        SLM & 34.6 & 247 & 26.8 & 465 & 28.3 & 295 & 60.8 & 197 & 37.6 & 301 & -- \\
        \quad w 1st SD & 34.6 & 199 & 26.8 & 351 & 28.3 & 248 & 60.8 & 193 & 37.6 & 248 & -- \\
        \hdashline
        QC$_\text{\appname}$ & 56.1 & 483 & 51.5 & 783 & 50.7 & 607 & 74.2 & 358 & 58.1 & 558 & 57.8 \\
        \quad wo IR & 55.0 & 485 & 50.8 & 780 & 50.4 & 606 & 73.0 & 358 & 57.3 & 557 & 55.6 \\
        \quad wo 1st SD & 56.1 & 602 & 51.5 & 1041 & 50.7 & 760 & 74.2 & 418 & 58.1 & 705 & 37.7 \\
        \quad wo 2nd SD & 56.1 & 510 & 51.5 & 843 & 50.7 & 623 & 74.2 & 374 & 58.1 & 588 & 52.2 \\
        \quad wo SD & 56.1 & 634 & 51.5 & 1112 & 50.7 & 783 & 74.2 & 435 & 58.1 & 741 & 34.7 \\
        \bottomrule
    \end{tabular}
    \end{threeparttable}
\end{table}

To investigate how effective each component of \appname to its effectiveness and efficiency, we conduct an ablation study using QwenCoder as the backbone model on the two benchmarks. 
We choose QwenCoder due to its high accuracy and fast inference speed observed in previous experiments. 
Specifically, we construct the following variants and compare them with \appname:
\begin{itemize}
    \item \textbf{w/o IR} removes the iterative retrieval component, i.e., only the initial retrieval results are used.
    \item \textbf{w 1st SD} enables the speculative decoding in the small model side.
    \item \textbf{w/o 1st SD} disables the speculative decoding in the small model side.
    \item \textbf{w/o 2nd SD} disables speculative decoding in the large model side.
    \item \textbf{w/o SD} disables the speculative decoding on both sides.
\end{itemize}

Since our target is striking a balance between effectiveness and efficiency, we additionally compute an \textit{Efficiency} metric, defined as the ratio between the percentage gain in accuracy and the percentage increase in latency over the SLM baseline. 
For better readability, we multiply this ratio by 100 when reporting results. 
Clearly, a higher value indicates a more favorable trade-off.

The experimental results are shown in Table~\ref{tab:ablation}.
We observe that removing any single component consistently reduces efficiency, demonstrating that each component contributes to the \appname.
In general, adopting iterative retrieval leads to an average performance improvement of 1.2\%, particularly achieving a 1.9\% improvement on the RepoEval-API (Left \& Right Context) dataset.
However, its benefit on RepoEval-Line is marginal. 
This is mainly due to the presence of samples that require completing import statements, where the small model's retrieval occasionally includes irrelevant content.
Since the time consumed by iterative retrieval is very short (with an average of 7 ms) and not all samples undergo iterative retrieval, removing it does not result in a significant change in latency.
Applying the Speculative Decoding to the SLM reduces the latency by 19.3\%. 
For the large model, incorporating two-stage speculative decoding substantially reduces latency.
Specifically, removing the 1st SR increases latency by 12.4\%–24.8\%, while removing the 2nd SD leads to a latency increase of 2.6\%-10.3\% across benchmarks. 
Removing both stages together leads to an even larger latency increase of 17.6\%–41.9\%.
This confirms that both SD stages contribute to faster generation by avoiding redundant token-by-token decoding.

\noindent \fbox{
    \parbox{0.95\linewidth}{\textbf{RQ3 Summary:}
    Both the 1st SD and 2nd SD are essential for latency reduction, and iterative retrieval provides valuable context. 
    Thus, both the two-stage speculative decoding and iterative retrieval yield measurable benefits, justifying their inclusion in the framework.}
}

\section{Discussion}\label{sec:Discussion}
\subsection{The Results of Routing Strategies and Cloud Model Invocation Reduction}


\begin{table*}[t]
\centering
\caption{The number of calls to the large model in different routing strategies}
\label{tab:api_call}
\begin{tabular}{lcccccccc}
\toprule
& \multicolumn{4}{c}{\textbf{Left \& Right Context}} & \multicolumn{4}{c}{\textbf{Only Left Context}} \\
 & \textbf{\makecell{RE\\(Line)}} & \textbf{\makecell{RE\\(API)}} & \textbf{\makecell{SE\\(Line)}} & \textbf{\makecell{SE\\(Suffix)}} & \textbf{\makecell{RE\\(Line)}} & \textbf{\makecell{RE\\(API)}} & \textbf{\makecell{SE\\(Line)}} & \textbf{\makecell{SE\\(Suffix)}} \\
\midrule
\#Samples & 1600 & 1600 & 999 & 999 & 1600 & 1600 & 999 & 999 \\
Static Routing & 248 & 202 & 128 & 95 & 614 & 619 & 523 & 207 \\
Dynamic Routing & 489 & 743 & 348 & 192 & 407 & 519 & 197 & 203 \\
Cloud Call & 737 & 945 & 476 & 287 & 1021 & 1138 & 720 & 410 \\
\bottomrule
\end{tabular}
\end{table*}

This section evaluates the impact of \appname’s routing strategies and assesses its impact on cloud-side computational resource savings.
Static Routing is designed to directly invoke the large model in cases where the local model exhibits low confidence in its prediction.
In contrast, dynamic routing calls the cloud model only when the user rejects the output of the local model; therefore, the number of invocations is determined by the local model’s accuracy.
Table~\ref{tab:api_call} reports the cloud model invocation counts under the two routing strategies in QC$_\text{\appname}$.
Overall, 39.0\% of the samples are routed to the cloud model under dynamic routing, implying that the local model achieves an accuracy of 61.0\%, which is substantially higher than its 46.4\% accuracy without static routing. 
This result shows that static routing effectively bypasses cases where the local model tends to fail.

The total number of large-model invocations under both strategies is summarized as the “Cloud Call” count in the table.
When both left and right contexts are available, \appname reduces cloud model calls by 40.9\%–71.3\%.
Under the left-context-only setting, static routing yields a smaller reduction since the local model performs less accurately in this scenario.
Nevertheless, \appname still lowers cloud model calls by 27.9\%–59.0\% in this setting.
These findings further highlight \appname’s effectiveness in reducing cloud-side computational overhead.

\rev{
\subsection{Reliability of Early Token Confidence}
\begin{table}[htbp]
  \centering
  \caption{Impact of token prefix length ($N$) on performance on the validation set}
  \label{tab:sensitivity_n}
  \begin{tabular}{lccccc}
    \toprule
    $N$ & 1 & 2 & 3 & 4 & 5 \\
    \midrule
    Precision & 0.41 & 0.51 & 0.53 & 0.53 & 0.54 \\
    Recall    & 0.91 & 0.86 & 0.90 & 0.92 & 0.93 \\
    F1        & 0.57 & 0.64 & 0.67 & 0.67 & 0.68 \\
    \bottomrule
  \end{tabular}
\end{table}
\begin{table}[ht]
\centering
\caption{Performance comparison between Average Probability and Joint Probability.}
\label{tab:AverageVSJoint}
\begin{tabular}{lrrrrrrrr}
\toprule
\multirow{2}{*}{Dataset} &
\multicolumn{4}{c}{Left \& Right Contexts} &
\multicolumn{4}{c}{Only Left Contexts} \\
\cmidrule(lr){2-5} \cmidrule(lr){6-9}
& \multicolumn{2}{c}{Latency (ms)} & \multicolumn{2}{c}{Accuracy (\%)} &
\multicolumn{2}{c}{Latency (ms)} & \multicolumn{2}{c}{Accuracy (\%)} \\
\cmidrule(lr){2-3} \cmidrule(lr){4-5} \cmidrule(lr){6-7} \cmidrule(lr){8-9}
& Avg. & Joint & Avg. & Joint & Avg. & Joint & Avg. & Joint \\
\midrule
\texttt{RepoEval-Line}   & 408 & \textbf{404} & \textbf{71.6} & 71.4 & 483 & \textbf{473} & \textbf{56.1} & 55.8 \\
\texttt{RepoEval-API}    & 719 & \textbf{700} & \textbf{65.3} & \textbf{65.3} & 783 & \textbf{759} & \textbf{51.5} & 51.3 \\
\texttt{StmtEval-Line}   & 495 & \textbf{486} & \textbf{71.2} & 70.5 & 607 & \textbf{578} & \textbf{50.7} & \textbf{50.7} \\
\texttt{StmtEval-Suffix} & \textbf{293} & 297 & \textbf{82.5} & 82.4 & 358 & \textbf{357} & \textbf{74.2} & 73.7 \\
\bottomrule
\end{tabular}
\end{table}
In the first part of our routing strategy, we use the average probability of the first $N$ tokens as a confidence score to decide whether to invoke the cloud model.
To justify the choice of $N=3$, we conducted a sensitivity analysis on the validation set described in Section~\ref{sec:training_set}, with results summarized in Table~\ref{tab:sensitivity_n}.
In this analysis, a sample is predicted positive if the average probability of the first $N$ tokens exceeds the threshold, while correctness is measured by exact match. 
The results show $N=1$ yields low Precision (0.41).
This is primarily because the first token in many Python statements is an indentation, which is trivial for language models to predict.
Consequently, relying solely on the first token leads to inflated confidence scores and numerous false positives.
Conversely, increasing $N$ to 4 or 5 offers negligible F1 gains (0.67 vs. 0.68)  but increases latency (approximately 18 ms per extra token). 
Therefore, $N=3$ represents a good reliability-efficiency trade-off.
We also compared two variants of the confidence metric: Average Probability versus Joint Probability (i.e., the product of token probabilities).
Fixing $N=3$ and using Qwen2.5-Coder-7B, we evaluated both metrics.
As shown in Table~\ref{tab:AverageVSJoint}, the performance difference is minimal: Joint Probability yields slightly lower latency, while Average Probability achieves marginally higher exact match accuracy.
Thus, it is acceptable to choose Average Probability.
}

\rev{
\subsection{Sparse vs. Dense Retrieval in Code Completion}
\begin{table}[htbp]
  \centering
  \caption{Performance comparison (Exact Match) between BM25 and Qwen-Embedding-0.6B.}
  \label{tab:retrieval_comparison}
  \begin{tabular}{lcccc}
    \toprule
    \multirow{2}{*}{Dataset} & \multicolumn{2}{c}{Prefix \& Suffix} & \multicolumn{2}{c}{Only Prefix} \\
    \cmidrule(lr){2-3} \cmidrule(lr){4-5}
     & BM25 & Qwen-Emb & BM25 & Qwen-Emb \\
    \midrule
    RepoEval-Line   & \textbf{66.5} & 65.0 & 51.8 & \textbf{52.4} \\
    RepoEval-API    & \textbf{60.1} & 59.5 & 48.3 & \textbf{48.6} \\
    StmtEval-Line   & \textbf{65.8} & 65.6 & \textbf{47.7} & \textbf{47.7} \\
    StmtEval-Suffix & \textbf{79.1} & 78.9 & 69.9 & \textbf{71.0} \\
    \bottomrule
  \end{tabular}
\end{table}
Our approach uses BM25 for context retrieval, as motivated in Section~\ref{sec:Approach}.
To further justify this decision, we compare BM25 with \texttt{Qwen-Embedding-0.6B}, a state-of-the-art embedding model for code. 
In the evaluation, the retrieved contexts from the two methods are fed into \texttt{Qwen2.5-Coder-7B} to generate completions.
As shown in Table~\ref{tab:retrieval_comparison}, the two methods perform comparably in exact match under different context settings.
However, BM25 incurs a latency of only 7 ms, whereas the dense retriever requires approximately 65 ms. 
This highlights BM25’s advantage in real-time code completion scenarios where efficiency is critical.
}

\subsection{Threats to validity}
The first threat involves the representativeness of our evaluation datasets.
Our used benchmarks may not perfectly reflect all real-world code completion scenarios. 
To mitigate this threat, we adopt a diverse set of test sets that cover a wide range of completion tasks, including line-level, API-level, and statement-level predictions, as well as completions from arbitrary positions. 
Additionally, we evaluate each benchmark under two configurations, i.e., with both prefix and suffix context, and with only prefix, to better simulate varying user behaviors in practical IDE usage.
Thus, we believe this threat is limited.
In future work, we plan to evaluate our framework on more benchmarks and real-world coding environments to further assess its generalizability and practical utility.

The second threat is the mismatch between our experimental setup and real-world deployment environments.
To mitigate this threat, we consulted practitioners from industry and learned that a typical deployment configuration involves hosting 7B-scale code models on A100 GPUs with an average network latency of approximately 200ms. 
We adopt the same latency assumption in our measurements. 
As we did not have access to A100 GPUs, we used A800 GPUs, which offer comparable performance, to conduct our experiments.
On the client side, we used an NVIDIA RTX 3080Ti to simulate the user's environment.
\rev{
This setup may be above the average hardware specifications found in developers' systems, posing a potential threat to the generalizability of our findings regarding client-side latency.
However, recent retail data from June 2025~\cite{GpuRetailSales} shows that top-selling GPUs outperform the 3080Ti, indicating that GPUs with performance exceeding 3080Ti are becoming increasingly accessible.
}
We plan to evaluate our approach across a broader range of client devices to better assess system performance under different hardware constraints.

The third threat is that our study focuses on Python, and thus the findings may not fully generalize to other programming languages.
However, Python is one of the most popular programming languages.
Furthermore, our framework is language-agnostic: all core components of our framework, including routing, speculative decoding, and iterative retrieval, operate without relying on any language-specific grammar rules or semantic knowledge.
As a result, it can be easily extended to other programming languages.
Another key factor preventing multilingual evaluation is the scarcity of high-quality local code completion models, coupled with the substantial cost and complexity of training such models.

The fourth threat is that our study focuses exclusively on line-level code completion, so the findings may not fully capture performance at larger granularities.
Nonetheless, our framework does not impose any restriction on completion granularity: it can be applied to larger-granularity tasks as long as suitable small models are available.
At the same time, our focus on line-level completion is well-aligned with real-world usage patterns, as mentioned in~\ref{subsec:benchmark}.

There is also a risk that our efficiency measurements may not accurately reflect actual inference speed due to potential resource contention or uncontrolled system factors. 
To mitigate this, we ensure that each GPU is dedicated to a single task during measurement and repeat each experiment five times to validate the consistency of results. 

\subsection{Limitation}
Although we have demonstrated the effectiveness of \appname through extensive experiments and analysis, its benefits in latency reduction and accuracy improvement may diminish if the small model is weak.
In such cases, users are more likely to reject the small model’s suggestions and fall back to the large model frequently, resulting in limited savings and additional overhead from running the small model.
However, our empirical results confirm that \appname is effective with an effective small model.
Moreover, as model compression techniques advance and the capabilities of small models continue to improve, we expect the performance of small models will improve continuously. 
At the same time, hardware on user devices is also becoming more powerful, making it feasible to deploy stronger small models locally.
Overall, we believe that our framework offers a meaningful and forward-looking direction for building more efficient code completion systems.

\section{Related Work}\label{sec:RelatedWork}
\subsection{Code completion}
Our work is closely related to recent advances in code completion research and tools.
Traditional studies~\cite{bruch2009learning, hou2010towards, asaduzzaman2014cscc} use rule-based methods or code examples for code completion.
Recently, numerous code completion tools have emerged in the market, such as Copilot~\cite{GitHubCopilotYour2025}, TabNine~\cite{TabnineAICode}, CodeWhisperer~\cite{AmazonCodeWhisperer}, and aiXcoder~\cite{jiangAiXcoder7BLightweightEffective2024}. 
These tools are predominantly based on cloud-hosted LLMs, which provide strong performance but can be slow and/or costly.
To address these limitations, a few tools have introduced local inference options~\cite{TongYiLingMa_NiDeZhiNengBianMaZhuShouALiYun, semenkin2025full}.
For example, FLCC~\cite{semenkin2025full} is a lightweight on-device model, enabling faster completions without relying on the cloud.
Nonetheless, they typically treat the local model as an isolated alternative rather than integrating it effectively with a larger model, which limits their performance.
In contrast, our study focuses on local–cloud collaboration, combining the low latency of local models with the high accuracy of cloud models through model cascading. 
We use static routing to assign tasks that the local model can handle quickly, reducing cloud calls, and dynamic routing guided by implicit user feedback to delegate difficult tasks to the cloud, improving completion quality. 
This combination balances efficiency and effectiveness, achieving both low-latency and high-accuracy code completions.

In addition to these commercial tools, leveraging repository context for code completion has been extensively explored in recent research~\cite{deng2025enhancing, ding2024cocomic, phan2024repohyper, zhang2023repocoder, wangRLCoderReinforcementLearning2024}. 
CoCoMIC~\cite{ding2024cocomic} improves retrieval accuracy through static analysis and infers with joint attention to in-file and retrieved cross-file context.
RepoHyper~\cite{phan2024repohyper} builds a graph representation of the project and leverages an expand-and-refine mechanism to improve the accuracy of code retrieval.
RepoCoder~\cite{zhang2023repocoder} leverages an iterative process of retrieval and generation to progressively refine code completions.
RLCoder~\cite{wangRLCoderReinforcementLearning2024} employs reinforcement learning to train an accurate retriever without labeled data and improve retrieval accuracy.
In contrast, our framework focuses on model cascading between small and large models, rather than enhancing a single model’s retrieval or generation.
These two directions are complementary: techniques developed for single-model improvements can be incorporated into our framework to further boost the performance of both the local and cloud models.


\subsection{Model Collaboration}
To improve performance while controlling computational cost, recent research has explored various strategies for collaborating multiple language models.
\textit{Cascade methods} reduce inference cost by invoking a sequence of models with increasing capacity, only escalating to more powerful models when simpler ones are insufficient~\cite{chen2023frugalgpt, chen2024modelCascading, fenggraphrouter, shnitzerLargeLanguageModel2023, chen2024octopus}.
Chen et al.~\cite{chen2024modelCascading} propose to generate unit tests alongside code completions, and use the pass/fail outcomes as a criterion to decide whether the current model’s output is sufficient or whether to escalate to a stronger model.
\textit{Model routing methods} focus on routing queries to different LLMs based on their specialization in specific domains~\cite{fenggraphrouter, shnitzerLargeLanguageModel2023, chen2024octopus, yu2025enhancing}.
For example, Octopus-v4~\cite{chen2024octopus} considers multiple LLMs with expertise in different domains and routes the queries to the one with the most matched topic.
\textit{Speculative decoding} employs a smaller model to generate draft tokens and a larger model to validate them in parallel~\cite{chen2024cascade, zhao2024ouroboros}.
For instance, CSDrafting~\cite{chen2024cascade} implements speculative decoding in a cascaded manner using multiple models, where the smallest model is a bigram language model combined with prompt lookup.

Compared to prior work, we are the first to adapt model collaboration to the local-cloud cascaded setting for line-level code completion. 
This scenario is highly interactive and places strong demands on responsiveness. 
Inspired by the interactive nature of this task, a novel contribution of our approach is the incorporation of implicit human feedback.
By leveraging this naturally occurring feedback loop, our approach effectively integrates real-time user input into the collaboration strategy, achieves a cost-free and reliable signal to guide model decisions.
Moreover, our approach introduces two key differences.
First, our method of leveraging contextual information through heuristic matching to generate high-quality drafts for the small model is non-trivial. 
Second, in contrast to cascades that ignore intermediate responses, our framework reuses rejected suggestions as retrieval cues, helping the large model generate higher-quality code completions.

\section{Conclusion}\label{sec:Conclusion}
In this work, we propose \appname, a novel code completion framework that effectively cascades the low-latency advantage of small local models with the high accuracy of large cloud-based models. 
By leveraging implicit user feedback, two-stage speculative decoding, and iterative retrieval, \appname substantially reduces the latency while delivering competitive or even superior accuracy, compared to LLM-based baselines.
Overall, our results highlight the practical value of cascading small and large models, and \appname offers better latency-accuracy trade-offs for interactive code completion.

\section{Data Availability}\label{sec:DataAvailability}
Our code and data are available at https://github.com/ZJU-CTAG/MCCom.

\begin{acks}
This research/project is supported by the National Natural Science Foundation of China (No.92582107), Zhejiang Provincial Natural Science Foundation of China (No.LZ25F020003), and Tencent Basic Platform Technology Rhino-Bird Focused Research Program.
\end{acks}

\balance
\bibliographystyle{ACM-Reference-Format}
\bibliography{cite}

@inproceedings{amann2016study,
  title={A study of visual studio usage in practice},
  author={Amann, Sven and Proksch, Sebastian and Nadi, Sarah and Mezini, Mira},
  booktitle={2016 IEEE 23rd International Conference on Software Analysis, Evolution, and Reengineering},
  volume={1},
  pages={124--134},
  year={2016},
  organization={IEEE}
}

@article{murphy2006java,
  title={How are Java software developers using the Eclipse IDE?},
  author={Murphy, Gail C and Kersten, Mik and Findlater, Leah},
  journal={IEEE software},
  volume={23},
  number={4},
  pages={76--83},
  year={2006}
}

@inproceedings{asaduzzaman2014cscc,
  title={Cscc: Simple, efficient, context sensitive code completion},
  author={Asaduzzaman, Muhammad and Roy, Chanchal K and Schneider, Kevin A and Hou, Daqing},
  booktitle={2014 IEEE International Conference on Software Maintenance and Evolution},
  pages={71--80},
  year={2014},
  organization={IEEE}
}

@inproceedings{bruch2009learning,
  title={Learning from examples to improve code completion systems},
  author={Bruch, Marcel and Monperrus, Martin and Mezini, Mira},
  booktitle={Proceedings of the 7th joint meeting of the European software engineering conference and the ACM SIGSOFT symposium on the foundations of software engineering},
  pages={213--222},
  year={2009}
}

@inproceedings{semenkin2025full,
  title={Full line code completion: Bringing ai to desktop},
  author={Semenkin, Anton and Bibaev, Vitaliy and Sokolov, Yaroslav and Krylov, Kirill and Kalina, Alexey and Khannanova, Anna and Savenkov, Danila and Rovdo, Darya and Davidenko, Igor and Karnaukhov, Kirill and others},
  booktitle={2025 IEEE/ACM 47th International Conference on Software Engineering: Software Engineering in Practice},
  pages={563--574},
  year={2025},
  organization={IEEE}
}

@inproceedings{leviathan2023fast,
  title={Fast inference from transformers via speculative decoding},
  author={Leviathan, Yaniv and Kalman, Matan and Matias, Yossi},
  booktitle={International Conference on Machine Learning},
  pages={19274--19286},
  year={2023},
  organization={PMLR}
}

@inproceedings{zhang2023repocoder,
  title={RepoCoder: Repository-Level Code Completion Through Iterative Retrieval and Generation},
  author={Zhang, Fengji and Chen, Bei and Zhang, Yue and Keung, Jacky and Liu, Jin and Zan, Daoguang and Mao, Yi and Lou, Jian-Guang and Chen, Weizhu},
  booktitle={Proceedings of the 2023 Conference on Empirical Methods in Natural Language Processing},
  pages={2471--2484},
  year={2023}
}

@online{GitHubCopilotYour2025,
  title = {{{GitHub Copilot}} · {{Your AI}} Pair Programmer},
  date = {2025},
  url = {https://github.com/features/copilot},
  urldate = {2025-05-06},
  langid = {english},
  organization = {GitHub}
}

@online{TabnineAICode,
  title = {Tabnine {{AI Code Assistant}} | Private, Personalized, Protected},
  url = {https://www.tabnine.com/},
  urldate = {2025-05-06},
  langid = {american},
  organization = {Tabnine}
}

@online{WindsurfFormerlyCodeium,
  title = {Windsurf (Formerly {{Codeium}}) - {{The}} Most Powerful {{AI Code Editor}}},
  url = {https://windsurf.com/},
  urldate = {2025-05-06}
}

@online{AmazonCodeWhisperer,
  title = {Amazon CodeWhisperer},
  url = {https://aws.amazon.com/ codewhisperer/},
  urldate = {2025-05-06}
}

@inproceedings{jiangAiXcoder7BLightweightEffective2024,
  title={aiXcoder-7B: A Lightweight and Effective Large Language Model for Code Processing},
  author={Jiang, Siyuan and Li, Jia and Zong, He and Liu, Huanyu and Zhu, Hao and Hu, Shukai and Li, Erlu and Ding, Jiazheng and Han, Yu and Ning, Wei and others},
  booktitle={2025 IEEE/ACM 47th International Conference on Software Engineering: Software Engineering in Practice},
  pages={215--226},
  year={2025},
  organization={IEEE}
}

@online{TongYiLingMa_NiDeZhiNengBianMaZhuShouALiYun,
  title = {TongYiLingMa},
  url = {https://lingma.aliyun.com/},
  urldate = {2025-05-06}
}

@article{lozhkov2024starcoder,
  title={Starcoder 2 and the stack v2: The next generation},
  author={Lozhkov, Anton and Li, Raymond and Allal, Loubna Ben and Cassano, Federico and Lamy-Poirier, Joel and Tazi, Nouamane and Tang, Ao and Pykhtar, Dmytro and Liu, Jiawei and Wei, Yuxiang and others},
  journal={arXiv preprint arXiv:2402.19173},
  year={2024}
}

@INPROCEEDINGS{UniGenCoder,
  author={Shao, Liangying and Yan, Yanfu and Poshyvanyk, Denys and Su, Jinsong},
  booktitle={2025 IEEE/ACM 47th International Conference on Software Engineering: New Ideas and Emerging Results (ICSE-NIER)}, 
  title={UniGenCoder: Merging SEQ2SEQ and SEQ2TREE Paradigms for Unified Code Generation}, 
  year={2025},
  volume={},
  number={},
  pages={71-75},
  doi={10.1109/ICSE-NIER66352.2025.00020}}

@inproceedings{fenggraphrouter,
  title={GraphRouter: A Graph-based Router for LLM Selections},
  author={Feng, Tao and Shen, Yanzhen and You, Jiaxuan},
  booktitle={The Thirteenth International Conference on Learning Representations},
  year={2025}
}

@article{shnitzerLargeLanguageModel2023,
  title={Large language model routing with benchmark datasets},
  author={Shnitzer, Tal and Ou, Anthony and Silva, M{\'\i}rian and Soule, Kate and Sun, Yuekai and Solomon, Justin and Thompson, Neil and Yurochkin, Mikhail},
  journal={arXiv preprint arXiv:2309.15789},
  year={2023}
}

@inproceedings{levenshtein1966binary,
  title={Binary codes capable of correcting deletions, insertions, and reversals},
  author={Levenshtein, Vladimir I and others},
  booktitle={Soviet physics doklady},
  volume={10},
  number={8},
  pages={707--710},
  year={1966},
  organization={Soviet Union}
}

@inproceedings{ding2024cocomic,
  title={Cocomic: Code completion by jointly modeling in-file and cross-file context},
  author={Ding, Yangruibo and Wang, Zijian and Ahmad, Wasi and Ramanathan, Murali Krishna and Nallapati, Ramesh and Bhatia, Parminder and Roth, Dan and Xiang, Bing},
  booktitle={Proceedings of the 2024 Joint International Conference on Computational Linguistics, Language Resources and Evaluation},
  pages={3433--3445},
  year={2024}
}

@inproceedings{wangRLCoderReinforcementLearning2024,
  title={RLCoder: Reinforcement Learning for Repository-Level Code Completion},
  author={Wang, Yanlin and Wang, Yanli and Guo, Daya and Chen, Jiachi and Zhang, Ruikai and Ma, Yuchi and Zheng, Zibin},
  booktitle={2025 IEEE/ACM 47th International Conference on Software Engineering},
  pages={165--177},
  year={2024},
  organization={IEEE Computer Society}
}

@article{eghbali2024hallucinator,
  title={De-hallucinator: Iterative grounding for llm-based code completion},
  author={Eghbali, Aryaz and Pradel, Michael},
  journal={arXiv preprint arXiv:2401.01701},
  year={2024}
}

@inproceedings{wuRepoformerSelectiveRetrieval2024,
  title={REPOFORMER: selective retrieval for repository-level code completion},
  author={Wu, Di and Ahmad, Wasi Uddin and Zhang, Dejiao and Ramanathan, Murali Krishna and Ma, Xiaofei},
  booktitle={Proceedings of the 41st International Conference on Machine Learning},
  pages={53270--53290},
  year={2024}
}

@article{huMiniCPMUnveilingPotential2024,
  title={Minicpm: Unveiling the potential of small language models with scalable training strategies},
  author={Hu, Shengding and Tu, Yuge and Han, Xu and He, Chaoqun and Cui, Ganqu and Long, Xiang and Zheng, Zhi and Fang, Yewei and Huang, Yuxiang and Zhao, Weilin and others},
  journal={arXiv preprint arXiv:2404.06395},
  year={2024}
}

@article{bavarianEfficientTrainingLanguage2022,
  title={Efficient training of language models to fill in the middle},
  author={Bavarian, Mohammad and Jun, Heewoo and Tezak, Nikolas and Schulman, John and McLeavey, Christine and Tworek, Jerry and Chen, Mark},
  journal={arXiv preprint arXiv:2207.14255},
  year={2022}
}

@inproceedings{huangOpenCoderOpenCookbook2024,
  title={Opencoder: The open cookbook for top-tier code large language models},
  author={Huang, Siming and Cheng, Tianhao and Liu, Jason Klein and Xu, Weidi and Hao, Jiaran and Song, Liuyihan and Xu, Yang and Yang, Jian and Liu, Jiaheng and Zhang, Chenchen and others},
  booktitle={Proceedings of the 63rd Annual Meeting of the Association for Computational Linguistics},
  pages={33167--33193},
  year={2025}
}

@article{wangPractitionersExpectationsCode2023,
  title={Practitioners' expectations on code completion},
  author={Wang, Chaozheng and Hu, Junhao and Gao, Cuiyun and Jin, Yu and Xie, Tao and Huang, Hailiang and Lei, Zhenyu and Deng, Yuetang},
  journal={arXiv preprint arXiv:2301.03846},
  year={2023}
}

@article{hui2024qwen2,
  title={Qwen2. 5-coder technical report},
  author={Hui, Binyuan and Yang, Jian and Cui, Zeyu and Yang, Jiaxi and Liu, Dayiheng and Zhang, Lei and Liu, Tianyu and Zhang, Jiajun and Yu, Bowen and Lu, Keming and others},
  journal={arXiv preprint arXiv:2409.12186},
  year={2024}
}

@article{guo2024deepseek,
  title={DeepSeek-Coder: When the Large Language Model Meets Programming--The Rise of Code Intelligence},
  author={Guo, Daya and Zhu, Qihao and Yang, Dejian and Xie, Zhenda and Dong, Kai and Zhang, Wentao and Chen, Guanting and Bi, Xiao and Wu, Yu and Li, YK and others},
  journal={arXiv preprint arXiv:2401.14196},
  year={2024}
}

@article{chen2023frugalgpt,
  title={FrugalGPT: How to Use Large Language Models While Reducing Cost and Improving Performance},
  author={Chen, Lingjiao and Zaharia, Matei and Zou, James},
  journal={Transactions on Machine Learning Research},
  year={2024}
}

@article{chen2024octopus,
  title={Octopus v4: Graph of language models},
  author={Chen, Wei and Li, Zhiyuan},
  journal={arXiv preprint arXiv:2404.19296},
  year={2024}
}

@article{phan2024repohyper,
  title={Repohyper: Better context retrieval is all you need for repository-level code completion},
  author={Phan, Huy Nhat and Phan, Hoang Nhat and Nguyen, Tien N and Bui, Nghi DQ},
  journal={CoRR},
  year={2024}
}

@inproceedings{hou2010towards,
  title={Towards a better code completion system by API grouping, filtering, and popularity-based ranking},
  author={Hou, Daqing and Pletcher, David M},
  booktitle={Proceedings of the 2nd International Workshop on Recommendation Systems for Software Engineering},
  pages={26--30},
  year={2010}
}

@article{roziere2023code,
  title={Code llama: Open foundation models for code},
  author={Roziere, Baptiste and Gehring, Jonas and Gloeckle, Fabian and Sootla, Sten and Gat, Itai and Tan, Xiaoqing Ellen and Adi, Yossi and Liu, Jingyu and Sauvestre, Romain and Remez, Tal and others},
  journal={arXiv preprint arXiv:2308.12950},
  year={2023}
}

@article{liu2024graphcoder,
  title={Graphcoder: Enhancing repository-level code completion via code context graph-based retrieval and language model},
  author={Liu, Wei and Yu, Ailun and Zan, Daoguang and Shen, Bo and Zhang, Wei and Zhao, Haiyan and Jin, Zhi and Wang, Qianxiang},
  journal={arXiv preprint arXiv:2406.07003},
  year={2024}
}

@article{rezaei2025vendi,
  title={Vendi-rag: Adaptively trading-off diversity and quality significantly improves retrieval augmented generation with llms},
  author={Rezaei, Mohammad Reza and Dieng, Adji Bousso},
  journal={arXiv preprint arXiv:2502.11228},
  year={2025}
}

@inproceedings{sagtani2025improving,
  title={Improving fim code completions via context \& curriculum based learning},
  author={Sagtani, Hitesh and Mehrotra, Rishabh and Liu, Beyang},
  booktitle={Proceedings of the Eighteenth ACM International Conference on Web Search and Data Mining},
  pages={801--810},
  year={2025}
}

@article{chen2024cascade,
  title={Cascade speculative drafting for even faster llm inference},
  author={Chen, Ziyi and Yang, Xiaocong and Lin, Jiacheng and Sun, Chenkai and Chang, Kevin and Huang, Jie},
  journal={Advances in Neural Information Processing Systems},
  volume={37},
  pages={86226--86242},
  year={2024}
}

@inproceedings{zhang2025coderag,
  title={CodeRAG: Finding Relevant and Necessary Knowledge for Retrieval-Augmented Repository-Level Code Completion},
  author={Zhang, Sheng and Ding, Yifan and Lian, Shuquan and Song, Shun and Li, Hui},
  booktitle={Proceedings of the 2025 Conference on Empirical Methods in Natural Language Processing},
  pages={23289--23299},
  year={2025}
}

@online{sombanerjeeGitHubCopilotChat2025,
  title = {{{GitHub Copilot Chat Explained}}: {{The Life}} of a {{Prompt}}},
  shorttitle = {{{GitHub Copilot Chat Explained}}},
  author = {{sombanerjee}},
  date = {2025-02-10T17:43:26+00:00},
  url = {https://devblogs.microsoft.com/all-things-azure/github-copilot-chat-explained-the-life-of-a-prompt/},
  urldate = {2025-09-10},
  langid = {american},
  organization = {All things Azure},
}

@article{zhao2024ouroboros,
  title={Ouroboros: Speculative decoding with large model enhanced drafting},
  author={Zhao, Weilin and Huang, Yuxiang and Han, Xu and Xiao, Chaojun and Liu, Zhiyuan and Sun, Maosong},
  journal={arXiv preprint arXiv:2402.13720},
  year={2024}
}

@inproceedings{chen2024modelCascading,
  title={Model cascading for code: A cascaded black-box multi-model framework for cost-efficient code completion with self-testing},
  author={Chen, Boyuan and Zhu, Mingzhi and Dolan-Gavitt, Brendan and Shafique, Muhammad and Garg, Siddharth},
  booktitle={2025 International Joint Conference on Neural Networks},
  pages={1--9},
  year={2025},
  organization={IEEE}
}

@book{glass2002facts,
  title={Facts and fallacies of software engineering},
  author={Glass, Robert L},
  year={2002},
  publisher={Addison-Wesley Professional}
}

@misc{GpuRetailSales,
  author = {TechEpiphany},
  title = {GPU Retail Sales June '25 Amazon US},
  howpublished = {\url{https://x.com/TechEpiphanyYT/status/1941064015016341736?s=20}},
  year = {2025},
  month = {7},
  day = {4}
}

@article{yu2025enhancing,
  title={Enhancing Domain-Specific Code Completion via Collaborative Inference with Large and Small Language Models},
  author={Yu, Jingrong and Gao, Zhipeng and Bao, Lingfeng and Liu, Zhongxin},
  journal={ACM Transactions on Software Engineering and Methodology},
  year={2025}
}

@article{deng2025enhancing,
  title={Enhancing project-specific code completion by inferring internal api information},
  author={Deng, Le and Ren, Xiaoxue and Ni, Chao and Liang, Ming and Lo, David and Liu, Zhongxin},
  journal={IEEE Transactions on Software Engineering},
  year={2025}
}

\end{document}